\documentclass[twocolumn,pra,showpacs,floatfix]{revtex4}
\usepackage{graphicx}
\usepackage{amssymb,amsmath,latexsym}
\usepackage{bm}
\usepackage[dvips]{epsfig}

\renewcommand{\vec}{\mathbf}
\newcommand{\ua}{\uparrow}
\newcommand{\da}{\downarrow}
\newcommand{\uda}{{\uparrow\downarrow}}
\newcommand{\dua}{{\downarrow\uparrow}}
\newcommand{\sech}{\mathrm{\,sech\,}}
\newcommand{\csch}{\mathrm{\,csch\,}}
\newcommand{\secha}{\mathrm{sech\,}}

\begin{document}

%%%%%%%%%%%%%%%%%%%%%%%%%%%%%%%%%%%%%%%%%%%%%%%%%%%%%%%%%%%%%%%%%%%%
\title{Effect of spin diffusion on current generated by spin motive force}
\author{Kyoung-Whan Kim$^1$, Jung-Hwan Moon$^2$, Kyung-Jin Lee$^2$, and Hyun-Woo Lee$^1$}
\affiliation{$^1$PCTP and Department of Physics, Pohang University of Science
and Technology, Pohang, 790-784, Korea\\$^2$Department of Materials Science
and Engineering, Korea University, Seoul 136-701, Korea}

\date{\today}

%%%%%%%%%%%%%%%%%%%%%%%%%%%%%%%%%%%%%%%%%%%%%%%%%%%%%%%%%%%%%%%%%%%%%

\begin{abstract}
Spin motive force is a spin-dependent force on conduction electrons induced
by magnetization dynamics. In order to examine its effects on magnetization
dynamics, it is indispensable to take into account spin accumulation, spin
diffusion, and spin-flip scattering since the spin motive force is in general
nonuniform. We examine the effects of all these on the way the spin motive
force generates the charge and spin currents in conventional situations,
where the conduction electron spin relaxation dynamics is much faster than
the magnetization dynamics. When the spin-dependent electric field is
spatially localized, which is common in experimental situations, we find that
the conservative part of the spin motive force is unable to generate the
charge current due to the cancelation effect by the diffusion current. We
also find that the spin current is a nonlocal function of the spin motive
force and can be effectively expressed in terms of nonlocal Gilbert damping
tensor. It turns out that any spin independent potential such as Coulomb
potential does not affect our principal results. At the last part of this
paper, we apply our theory to current-induced domain wall motion.
\end{abstract}

%\pacs{}

\maketitle
%%%%%%%%%%%%%%%%%%%%%%%%%%%%%%%%%%%%%%%%%%%%%%%%%%%%%%%%%%%%%%%%%%%%%%%%%%

\section{Introduction \label{sec_intro}}
In a ferromagnetic system, dynamics of space-time dependent magnetization
vector $\vec{M}(\vec{x},t)$ is described by the following phenomenological
equation \cite{Zhang04PRL,Thiaville05EPL,Tatara06JPS,Stiles07PRB}
\begin{eqnarray}
\frac{\partial\vec{M}}{\partial t}&=&-\gamma\vec{M}\times\vec{H}_{eff}+\frac{\alpha}{M_s}\vec{M}\times\frac{\partial\vec{M}}{\partial t}\nonumber\\
&&+\frac{\mu_B}{eM_s}(\vec{j}_s\cdot\nabla)\vec{M}-\frac{\beta\mu_B}{eM_s^2}\vec{M}\times(\vec{j}_s\cdot\nabla)\vec{M}. \label{LLG}
\end{eqnarray}
This is the Landau-Lifshitz-Gilbert (LLG) equation generalized to include the
spin-transfer torque terms (last two terms). Here, $\vec{H}_{eff}$ is the
functional derivative of energy with respect to $\vec{M}$, $\alpha$ is
Gilbert damping constant, $M_s$ is saturation magnetization, $\mu_B$ is Bohr
magneton, $e$ is electron charge, $\beta$ is nonadiabaticity, and $\vec{j}_s$
is spin polarized electric current given by difference between current by
spin-up and spin-down electrons. LLG equation describes the dynamics of
magnetization under applied electromagnetic fields. On the other hand, there
exists a reciprocal process; temporal and spatial variation of magnetization
induces additional electromagnetic fields on conduction electrons. These
fields are spin-dependent and in general nonconservative. Thus the resulting
spin-dependent motive force is called spin motive force (SMF). SMF is firstly
predicted by Berger \cite{Berger86PRB} and recently formulated by
generalizing Faraday's law \cite{Barnes07PRL}. It is also suggested
\cite{Zhang09PRL} that SMF is also described by spin pumping effect
\cite{Tserkovnyak02PRL}. Recently, SMF and its effect is intensively studied
\cite{Shibata09PRL,Zhang10PRB,Kim11Preprint,Zhang09PRL,Duine09PRB,Ohe09APL,Ohe09JAP,Tserkovnyak08PRB,Tserkovnyak09PRB,Wong10PRB,Hai09Nature,Yang09PRL,Saslow07PRB}
in this field.

Without any other perturbation, the explicit expressions of the induced spin
electromagnetic fields are known as
\cite{Shibata09PRL,Tserkovnyak09PRB,Tserkovnyak08PRB,Zhang09PRL,Zhang10PRB,Volovik87JPC,Kim11Preprint}
\begin{eqnarray}
E_{i}^{\uda}&=&\pm\frac{\hbar}{2eM_s^3}\left(\partial_t\vec{M}\times\partial_i\vec{M}\right)\cdot\vec{M},\label{E0}\\
B_{i}^{\uda}&=&\mp\frac{\hbar}{2eM_s^3}\epsilon_{ijk}(\partial_j\vec{M}\times\partial_k\vec{M})\cdot\vec{M},\label{B0}
\end{eqnarray}
where $\ua$ and $\da$ stand for spin-up and down electrons. The fields
generate spin-dependent Lorentz force
$F_s^{\uda}=e(\vec{E}_i^{\uda}+\vec{v}^{\uda}\times\vec{B}_s^{\uda})$. While
the term ``motive force" refers to quantities of voltage dimension, the term
SMF is sometimes used to denote $F_s^\uda$. In this paper, we adopt the
latter terminology. It is easily noticed that SMF is spin dependent,
nonconservative, spatially varying, and localized in general situations. As
one can see from Eqs. (\ref{E0}) and (\ref{B0}), SMF is usually too small to
be measured directly. Another way is to study the effect of SMF on
magnetization dynamics. Since SMF induces additional spin current, additional
spin-transfer torque arises and it changes LLG equation. Consequently, LLG
equation (without applied spin current) is modified as \cite{Zhang09PRL}
\begin{equation}
\frac{\partial\vec{M}}{\partial t}=-\gamma\vec{M}\times\vec{H}_{eff}+\frac{1}{M_s}\vec{M}\times\left(\mathcal{D}\cdot\frac{\partial\vec{M}}{\partial t}\right),\label{LLG(Z&Z)}
\end{equation}
where $\mathcal{D}$ is modified damping tensor given by
\begin{equation}
\mathcal{D}_{ij}=\alpha\delta_{ij}+\frac{\eta}{M_s^4}\sum_k(\vec{M}\times\partial_k\vec{M})_i(\vec{M}\times\partial_k\vec{M})_j,\label{damping(Z&Z)}
\end{equation}
and $\eta=\mu_B\hbar \sigma/2e^2M_s$. Here, $\sigma$ is electrical
conductivity.

However, the previous work has a crucial limitation that spin density has
been considered to be constant while spin density in reality is nonuniform
since SMF is spatially varying. The main consequence of the nonuniform spin
density $n^\uda(\vec{x},t)$ is diffusion current proportional to $-\nabla
n^\uda$, which suppresses the effect of SMF. Therefore, a realistic model
should take into account spin accumulation (nonuniform spin density),
diffusion and spin-flip scattering. The purpose of this paper is to find the
spin density $n^\uda$, diffusion current, total current induced by SMF, and
their effect on magnetization dynamics in the presence of spin accumulation,
diffusion, and spin-flip scattering. As one shall see in Sec.
\ref{sec_solution}, the solution of $n^\uda$ in the most general situation is
too complicated to study the effect on magnetization dynamics. To obtain
simple analytic expressions, we take an approximation that spin-flip time is
much shorter than the time scale of magnetization dynamics. As a final
comment, our result does not assume any specific form of SMF. Thus, it
remains valid for the modified SMF due to, for instance, nonadiabaticity
\cite{Tserkovnyak08PRB}, spin-orbit coupling \cite{Kim11Preprint}, and other
kinds of spin dependent electric field \cite{Sherman06arXiv}.

Several previous works are closely related to our work. Spin drift-diffusion
equation, which has similar form to our theory is suggested in Ref.
\cite{Tserkovnyak08PRB}. And, the effect on spin and charge current is
investigated from Boltzmann equation in Ref. \cite{Zhang10PRB}. We set our
starting point as the equation of motion of conduction electrons in Ref.
\cite{Zhang04PRL} to make our analysis consistent with previous theories in
this field. Different from the previous theories focusing on 1D, we
successfully generalized our result to 3D, and found that nonconservative
part of SMF plays a crucial role in current in a higher dimensional system.
In Sec. \ref{sec_solution}, we compare our result with the previous theory
qualitatively and quantitatively. In addition, we investigated the effect of
charge neutrality on our results. It turns out that charge neutrality
potential does not change our principal results, charge current and spin
current, even though it changes charge density and spin density. Furthermore,
we show that any spin independent potential cannot alter our principal
results, either.

This paper is organized as follows. In Sec. \ref{sec_sdde}, we construct the
spin drift-diffusion equation and introduce variables. Then, we solve the
equation in Sec. \ref{sec_solution}, and discuss various implications. In
Sec. \ref{sec_dwmotion}, we apply our result to current-induced domain wall
(DW) motion and briefly discuss the effect of spin diffusion. In Sec.
\ref{sec_side}, we generalize our theory for general boundary condition and
for general spin indendendent potentials. Finally, in Sec.
\ref{sec_conclusion}, there are concluding remarks. Technical details are in
Appendices.

\section{Model \label{sec_sdde}}
\subsection{Spin drift-diffusion equation}
To construct the equation of $n^\uda(\vec{x},t)$, we take the starting point
as the equation of spin density $\vec{m}$ of conduction electrons
\cite{Zhang04PRL},
\begin{equation}
\frac{\partial\vec{m}}{\partial t}+\nabla\cdot J=-\frac{1}{\tau_{ex}M_s}\vec{m}\times\vec{M}-\langle\vec{\Gamma}\rangle.
\label{contiuity eq}
\end{equation}
Here, $J$ is spin current tensor, $\tau_{ex}=\hbar/SJ_{ex}$, and $S$ denotes
the magnitude of spin of local magnetization. The left-hand side is based on
the continuity equation. The first term on the right-hand side is the
precession term due to the exchange coupling between conduction electrons and
magnetization. $\langle\vec{\Gamma}\rangle$ includes the effect of spin
scattering processes. Here, the second rank tensor $J$ is defined by
\begin{equation}
J=-\frac{\mu_B}{e}\vec{j}_s\otimes\frac{\vec{m}}{|\vec{m}|}.
\end{equation}
The effect of the perpendicular component to $\vec{M}$ of Eq. (\ref{contiuity
eq}) is already investigated by Zhang and Li \cite{Zhang04PRL}, and they
found the nonadiabatic term of LLG equation. In the absence of spin
accumulation, the magnitude of $\vec{m}$ is constant, so it suffices to solve
only the perpendicular component of the equation. However, in the presence of
the spin accumulation, the magnitude variation of $\vec{m}$ should be also
studied. We define spin number density $n_s\equiv|\vec{m}|/\mu_B$. Taking
care of the fact that $n_s$ has space-time dependence, the parallel component
of Eq. (\ref{contiuity eq}) to $\hat{m}\equiv\vec{m}/|\vec{m}|$ results in
\begin{equation}
\frac{\partial n_s}{\partial t}+\frac{1}{\mu_B}\hat{m}\cdot\langle\vec{\Gamma}\rangle=\frac{1}{e}\nabla\cdot\vec{j}_s.
\label{contiuity eq2}
\end{equation}
It is convenient to separate the variables to that of up and down electrons.
$n_s=n^\ua-n^\da$ and $\vec{j}_s=\vec{j}^\ua-\vec{j}^\da$. Here, $n^\uda$ and
$\vec{j}^\uda$ denote spin number density of spin-up/down electrons and
charge current density generated by spin-up/down electrons, respectively.
Equation (\ref{contiuity eq2}) is nothing but the continuity equation
containing spin nonconserving processes described by $\vec{\Gamma}$. To
obtain independent equations of spin-up/down electrons, we use the following
continuity equation of total electron number density
\begin{equation}
\frac{\partial n_e}{\partial t}=\frac{1}{e}\nabla\cdot\vec{j}_e,
\label{contiuity eq3}
\end{equation}
where $n_e=n^\ua+n^\da$ is electron number density and
$\vec{j}_e=\vec{j}^\ua+\vec{j}^\da$ is charge current density. Combining Eqs.
(\ref{contiuity eq2}) and (\ref{contiuity eq3}), one obtains
\begin{equation}
\frac{\partial n^\uda}{\partial t}\pm\frac{1}{2\mu_B}\hat{m}\cdot\langle\vec{\Gamma}\rangle=\frac{1}{e}\nabla\cdot\vec{j}^\uda.
\end{equation}
Note that $\vec{\Gamma}$ represents spin-flip scattering processes. As a
simple model, we take the well-known form of spin-flip scattering,
\begin{equation}
\frac{1}{2\mu_B}\hat{m}\cdot\langle\vec{\Gamma}\rangle=\frac{n^\ua}{\tau^\ua}-\frac{n^\da}{\tau^\da},
\label{spin-flip sc}
\end{equation}
where $\tau^\ua$ is characteristic time of the spin-flip scattering process
from spin-up to -down state, and $\tau^\da$ is similarly defined. Then,
\begin{equation}
\frac{\partial n^\uda}{\partial t}+\frac{n^\uda}{\tau^\uda}-\frac{n^\dua}{\tau^\dua}=\frac{1}{e}\nabla\cdot\vec{j}^\uda,
\label{sdde1}
\end{equation}
which is the spin drift-diffusion equation. Similar form of Eq. (\ref{sdde1})
was also suggested in Ref. \cite{Tserkovnyak08PRB}.

For simplicity, we may assume without losing generality that the SMF is
turned on at $t=0$ and that, for $t<0$, the system is in equilibrium. We set
$n^\uda(\vec{x},t=0)=n_0^\uda$, where $n_0^\uda$ is equilibrium electron
density of spin up and down at $t<0$. By definition, the equilibrium density
$n_0^\uda$ is the equilibrium solution of Eq. (12) for $t<0$. Inserting
$n^\uda(\vec{x},t<0)=n_0^\uda$ to Eq. (12), one obtains an important
constraint $n_0^\ua/\tau^\ua=n_0^\da/\tau^\da$. With the help of this
constraint, four variables $n_0^\ua$, $n_0^\da$, $\tau_\ua$ and $\tau^\da$
can be described by three variables, $n_0^\ua$, $n_0^\da$, and $\tau_{sf}$
($\tau_{sf}^{-1}=\tau^{\ua-1}+\tau^{\da-1}$). Then, Eq. (\ref{sdde1}) is
rewritten with only one spin-flip time $\tau_{sf}$. In addition, current
$\vec{j}^\uda$ can be written as $\sigma^\uda \vec{E}_s^\uda+eD^\uda\nabla
n^\uda$, where $\sigma^\uda$ and $\vec{E}_s^\uda$ are respectively the
conductivity and SMF (divided by $e$) for spin-up/down electrons. Then, one
straightforwardly obtains the final form of the equation of our model.
\begin{eqnarray}
&&\frac{\partial n^\uda}{\partial t}-D^\uda\nabla^2 n^\uda +\frac{n_0^\ua n_0^\da}{n_0^\ua+n_0^\da}\frac{1}{\tau_{sf}}
\left(\frac{n^\uda}{n_0^\uda}-\frac{n^\dua}{n_0^\dua}\right)\nonumber\\&&=\frac{\sigma^\uda}{e}\nabla\cdot \vec{E}_s^\uda.
\label{sdde2}
\end{eqnarray}
As mentioned in Sec. \ref{sec_intro}, we treat $\vec{E}_s^\uda$ as
nonconservative, spatially varying fields. In addition, it is assumed that
spin dependence of $\vec{E}_s$ is given by
$\vec{E}_s^\ua=-\vec{E}_s^\da\equiv\vec{E}_s$. Slight generalization of our
theory at the final step allows to investigate the formula for
$\vec{E}_s^\ua\ne-\vec{E}_s^\da$. No other restriction of $\vec{E}_s^\uda$ is
not assumed in order to obtain maximally generalized result.

As suggested in Ref. \cite{Tserkovnyak08PRB}, in realistic systems, the
Coulomb interaction should be taken into account. Hence, one introduces
Coulomb potential $V_c$ and add it to the spin motive force as
$\vec{E}_s^\uda\rightarrow \vec{E}_s^\uda-\nabla V_c$. The Coulomb
interaction strongly suppresses the charge accumulation. Mathematically the
interaction may thus be handled by imposing the charge neutrality constraint.
We show in Sec. \ref{sec_side} that charge neutrality constraint changes
electron densities, but not currents. Hence, the LLG equation is hardly
affected by charge neutrality potential. For this reason, we do not take into
account the Coulomb interaction until Sec. \ref{sec_side} in order to show
simple logical flow.

Note that all variables in Eq. (\ref{sdde2}) are not independent. Einstein's
relation is given by $\sigma^\uda=e^2 D^\uda N^\uda$ where $N^\uda$ is
density of states of spin-up/down electrons at Fermi energy. Since
$N^\uda\propto n_0^\uda$, one obtains $\sigma^\ua/D^\ua
n_0^\ua=\sigma^\da/D^\da n_0^\da$. This is one of the key constraints of our
model.

The solution of Eq. (\ref{sdde2}) is very complicated as one shall see in
Sec. \ref{sec_solution}. To gain an insight, it is illustrative to assume
that $\tau_{sf}$ is much smaller than the time scale of magnetization
dynamics so $\vec{E}_s$ is almost constant in time scale within $\tau_{sf}$.
We found that, in this limit, the solution is much simpler and it is easier
to catch physical meanings.

\subsection{Variable definitions and relations}

In Sec. \ref{sec_solution}, there appear several variables and quantities
which have not been defined yet. To help readers, we present definitions of
them here, rather than Sec. \ref{sec_solution}.

Since Eq. (\ref{sdde2}) is coupled, it is convenient to solve it in matrix
form. Hence, we define spin accumulation vector, which is a column vector
defined by
\begin{equation}
\mathcal{N}=\left(
             \begin{array}{c}
               n^\ua \\
               n^\da \\
             \end{array}
           \right).\label{NN}
\end{equation}
Similarly, we define current density and SMF vector.
\begin{eqnarray}
\mathcal{J}&=&\left(
                \begin{array}{c}
                  \vec{j}^\ua \\
                  \vec{j}^\da \\
                \end{array}
              \right),\\
\mathcal{E}&=&\left(
                \begin{array}{c}
                  \vec{E}_s^\ua \\
                  \vec{E}_s^\da \\
                \end{array}
              \right)=\vec{E}_s\left(
                \begin{array}{c}
                  1 \\
                  -1 \\
                \end{array}
              \right).\label{EE}
\end{eqnarray}
Equations (\ref{NN})-(\ref{EE}) are related by the following relation.
\begin{equation}
\mathcal{J}=\left(
              \begin{array}{cc}
                \sigma^\ua & 0 \\
                0 & \sigma^\da \\
              \end{array}
            \right)\mathcal{E}+e\left(
                                  \begin{array}{cc}
                                    D^\ua & 0 \\
                                    0 & D^\da \\
                                  \end{array}
                                \right)\nabla\mathcal{N}.\label{JJ}
\end{equation}
The first term in right-hand side corresponds to conventional electrical
current and the second term corresponds to diffusion current.

Instead of diffusion constants, it is more physical and intuitive to express
results in terms of spin-flip length which is defined by
$\lambda_{sf}^\uda=\sqrt{D^\uda\tau_{sf}}$. The averaged spin diffusion
length is also defined by the conventional way
\begin{equation}
\lambda_{sf}^{-2}=\frac{n_0^\ua\lambda_{sf}^{\ua2}+n_0^\da\lambda_{sf}^{\da2}}{(n_0^\ua+n_0^\da)\lambda_{sf}^{\ua2}\lambda_{sf}^{\da2}}.\label{lsf}
\end{equation}
By Einstein's relation, Eq. (\ref{lsf}) is equivalent to
\begin{equation}
\lambda_{sf}^{2}=\frac{\sigma^\ua\lambda_{sf}^{\da2}+\sigma^\da\lambda_{sf}^{\ua2}}{\sigma^\ua+\sigma^\da}.\label{lsf2}
\end{equation}
Combining Eqs. (\ref{lsf}) and (\ref{lsf2}), $\lambda_{sf}^\uda$ is
represented in terms of $\lambda_{sf}$.
\begin{equation}
\lambda_{sf}^{\uda2}=\lambda_{sf}^2\frac{\sigma}{\sigma^\dua}\frac{n_0^\dua}{n_0^\ua+n_0^\da},
\end{equation}
where $\sigma=\sigma^\ua+\sigma^\da$ is total electrical conductivity.

Conductivity polarization $P$ and density polarization $P_n$ are defined by
\begin{eqnarray}
P&=&\frac{\sigma^\ua-\sigma^\da}{\sigma},\\
P_n&=&\frac{n_0^\ua-n_0^\da}{n_0^\ua+n_0^\da}.
\end{eqnarray}
With these polarizations, $\sigma^\uda$ and $n_0^\uda$ are represented in
terms of $\sigma$ and $(n_0^\ua+n_0^\da)$ as $\sigma^\uda=(1\pm P)\sigma/2$
and $n_0^\uda=(1\pm P_n)(n_0^\ua+n_0^\da)/2$.

Lastly, we use a mathematical convention that $\tilde{A}(\vec{k})$ is the
Fourier transform of a position dependent function $A(\vec{x})$ with respect
to $\vec{x}$. That is,
\begin{eqnarray}
\tilde{A}(\vec{k})\equiv\mathcal{F}[A(\vec{x})](\vec{k})=\frac{1}{(2\pi)^{d/2}}\int A(\vec{x}) e^{-i\vec{k}\cdot\vec{x}}d^dx,
\end{eqnarray}
for a $d$-dimensional system.

\section{Charge and spin currents in the presence of spin diffusion \label{sec_solution}}
\subsection{Solution of the spin drift-diffusion equation for localized electric field}

Before solving Eq. (\ref{sdde2}) for general cases, we first solve the
equation for localized $\vec{E}_s$ since $\vec{E}_s$ is localized in most
cases. In Sec. \ref{sec_side}, we generalize our theory to include spatially
extended $\vec{E}_s$.

Since $\vec{E}_s$ is localized, it is possible to take Fourier transform with
respect to position. Then, $\tilde{\mathcal{E}}$ and $\tilde{\mathcal{N}}$
are well-defined localized functions of $\vec{k}$ except initial condition
part. In addition, localized $\vec{E}_s$ implies that the boundary condition
is given by $n^\uda(|\vec{x}|\rightarrow\infty,t)=n_0^\uda$ because
$\vec{E}_s(|\vec{x}|\rightarrow\infty,t)=0$ does not affect spin density.
After Fourier transform, Eq. (\ref{sdde2}) is written as, in matrix form,
\begin{equation}
\frac{\partial \tilde{\mathcal{N}}}{\partial t}+\frac{\Omega}{\tau_{sf}}\tilde{\mathcal{N}}
=\frac{1}{e}\left(
   \begin{array}{cc}
     \sigma^\ua & 0 \\
     0 & \sigma^\da \\
   \end{array}
 \right)i\vec{k}\cdot\tilde{\mathcal{E}},\label{sdde3}
\end{equation}
where
\begin{equation}
\Omega=\left(
         \begin{array}{cc}
           \lambda_{sf}^{\ua2}k^2+\frac{n_0^\da}{n_0^\ua+n_0^\da} & -\frac{n_0^\ua}{n_0^\ua+n_0^\da} \\
           -\frac{n_0^\da}{n_0^\ua+n_0^\da} & \lambda_{sf}^{\da2}k^2+\frac{n_0^\ua}{n_0^\ua+n_0^\da} \\
         \end{array}
       \right).\label{omega}
\end{equation}
Equation (\ref{sdde3}) is a first order ordinary differential equation with
respect to $t$ and the initial condition is given by
$\tilde{\mathcal{N}}(\vec{k},t=0)=(2\pi)^{d/2}\delta(\vec{k})(n_0^\ua~n_0^\da)^T$.
The solution is simply given by
\begin{eqnarray}
\tilde{\mathcal{N}}(\vec{k},t)&=&e^{-\Omega t/\tau_{sf}}\tilde{\mathcal{N}}(\vec{k},0)\label{solution N}\\
&&+\int_0^t e^{-\Omega(t-t')/\tau_{sf}}\left(
                                         \begin{array}{cc}
                                           \frac{\sigma^\ua}{e} & 0 \\
                                           0 & \frac{\sigma^\da}{e} \\
                                         \end{array}
                                       \right)i\vec{k}\cdot\tilde{\mathcal{E}}(\vec{k},t')dt'.\nonumber
\end{eqnarray}
Since the first term of Eq. (\ref{solution N}) represents the time variation
of equilibrium number density, one can realize that the term should be given
by $e^{-\Omega
t/\tau_{sf}}\tilde{\mathcal{N}}(\vec{k},0)=\tilde{\mathcal{N}}(\vec{k},0)$.
Mathematical derivation of this argument is given in Appendix
\ref{sec_appen_n0}. The second term of Eq. (\ref{solution N}) is almost
impossible to take inverse Fourier transform. Hence, as mentioned, we use an
approximation that $\tau_{sf}$ is very small. In this limit, Appendix
\ref{sec_appen_delta} shows that
\begin{equation}
e^{-\Omega t/\tau_{sf}}\Theta(t)\approx\tau_{sf}\Omega^{-1}\delta(t),
\label{delta}
\end{equation}
where $\Theta(t)$ is Heaviside step function. By this approximation, solution
of the spin drift-diffusion equation Eq. (\ref{solution N}) becomes
\begin{equation}
\tilde{\mathcal{N}}(\vec{k},t)=\tilde{\mathcal{N}}(\vec{k},0)+\frac{\tau_{sf}}{e}\Omega^{-1}
\left(
  \begin{array}{cc}
    \sigma^\ua & 0 \\
    0 & \sigma^\da \\
  \end{array}
\right)i\vec{k}\cdot\tilde{\mathcal{E}}(\vec{k},t).\label{solution N2}
\end{equation}
The inverse of $\Omega$ is explicitly given by
\begin{eqnarray}
\Omega^{-1}&=&\frac{1}{\det\Omega}\left(
                                    \begin{array}{cc}
                                      \lambda_{sf}^{\da2}k^2+\frac{n_0^\ua}{n_0^\ua+n_0^\da} & \frac{n_0^\ua}{n_0^\ua+n_0^\da} \\
                                      \frac{n_0^\da}{n_0^\ua+n_0^\da} & \lambda_{sf}^{\ua2}k^2+\frac{n_0^\da}{n_0^\ua+n_0^\da} \\
                                    \end{array}
                                  \right),\label{omega^-1}\\
\det\Omega&=&\lambda_{sf}^{\ua2}\lambda_{sf}^{\da2}k^2\left(k^2+\lambda_{sf}^{-2}\right).
\end{eqnarray}
Now, excited charge density $\Delta n_e\equiv
(n^\ua+n^\da)-(n_0^\ua+n_0^\da)$ and excited spin density $\Delta
n_s\equiv(n^\ua-n^\da)-(n_0^\ua-n_0^\da)$ are given by,
\begin{eqnarray}
\Delta\tilde{n}_e(\vec{k},t)&=&\left(\begin{array}{cc}
                           1 & 1
                         \end{array}\right)\tilde{\mathcal{N}}(\vec{k},t)-(2\pi)^{d/2}\delta(\vec{k})(n_0^\ua+n_0^\da)\nonumber\\
                      &=&\frac{\sigma\tau_{sf}}{e\lambda_{sf}^2}\frac{1-P^2}{1-P_n^2}\frac{P_nk^2+P\lambda_{sf}^{-2}}{k^2+\lambda_{sf}^{-2}}\frac{i\vec{k}\cdot\tilde{\vec{E}}_s(\vec{k},t)}{k^2},\label{Delta ne}\\
\Delta\tilde{n}_s(\vec{k},t)&=&\left(\begin{array}{cc}
                           1 & -1
                         \end{array}\right)\tilde{\mathcal{N}}(\vec{k},t)-(2\pi)^{d/2}\delta(\vec{k})(n_0^\ua-n_0^\da)\nonumber\\
                      &=&\frac{\sigma\tau_{sf}}{e\lambda_{sf}^2}\frac{1-P^2}{1-P_n^2}\frac{k^2+P_nP\lambda_{sf}^{-2}}{k^2+\lambda_{sf}^{-2}}\frac{i\vec{k}\cdot\tilde{\vec{E}}_s(\vec{k},t)}{k^2}
\end{eqnarray}
in $\vec{k}$-space.

At this stage, there is no need to show complicated real space expressions of
the densities, because what affects LLG equation mainly is spin current. In
the next subsection, we find the expressions of charge and spin currents in
bith $k$-space and real space.

\subsection{Charge and spin currents}

Charge and spin currents in the absence of spin diffusion are given by
\begin{eqnarray}
\vec{j}_e&=&\sigma^\ua\vec{E}_s^\ua+\sigma^\da\vec{E}_s^\ua=P\sigma \vec{E}_{s},\label{je0}\\
\vec{j}_s&=&\sigma^\ua\vec{E}_s^\ua-\sigma^\da\vec{E}_s^\ua=\sigma \vec{E}_{s}.\label{js0}
\end{eqnarray}
In this subsection, how the spin current and charge current generated by SMF
is changed by spin diffusion from Eqs. (\ref{je0}) and (\ref{js0}) is
examined with the help of Eqs. (\ref{JJ}) and (\ref{solution N2}). After some
algebra,
\begin{eqnarray}
\tilde{\mathcal{J}}&=&\left(
              \begin{array}{cc}
                \sigma^\ua & 0 \\
                0 & \sigma^\da \\
              \end{array}
            \right)\tilde{\mathcal{E}}+e\left(
                                  \begin{array}{cc}
                                    D^\ua & 0 \\
                                    0 & D^\da \\
                                  \end{array}
                                \right)i\vec{k}\tilde{\mathcal{N}}\nonumber\\
            &=&\tilde{\vec{E}}_s\nonumber\left(\begin{array}{c}\sigma^\ua \\ -\sigma^\da\end{array}\right)\\&&
            +\frac{i\vec{k}(i\vec{k}\cdot\tilde{\vec{E}}_s)}{k^2\left(k^2+\lambda_{sf}^{-2}\right)}
            \left(\begin{array}{c}\sigma^\ua k^2+P\sigma^\ua\lambda_{sf}^{-2} \\ -\sigma^\da k^2+P\sigma^\da\lambda_{sf}^{-2}\end{array}\right).
\end{eqnarray}
Now, the expressions of charge current
$\tilde{\vec{j}}_e=\tilde{\vec{j}}^\ua+\tilde{\vec{j}}^\da$ and spin current
$\tilde{\vec{j}}_s=\tilde{\vec{j}}^\ua-\tilde{\vec{j}}^\da$ are
straightforward.
\begin{eqnarray}
\tilde{\vec{j}}_e&=&P\sigma\left[\tilde{\vec{E}}_s+\frac{i\vec{k}(i\vec{k}\cdot\tilde{\vec{E}}_s)}{k^2}\right],\label{je(k)}\\
\tilde{\vec{j}}_s&=&P\tilde{\vec{j}}_e+(1-P^2)\sigma\left(\tilde{\vec{E}}_s+\frac{i\vec{k}(i\vec{k}\cdot\tilde{\vec{E}}_s)}{k^2+\lambda_{sf}^{-2}}\right).\label{js(k)}
\end{eqnarray}
We present $\vec{j}_s$ in terms of $\vec{j}_e$ in order for one to see easily
$\vec{j}_s=\vec{j}_e$ for $P=\pm1$ ; for perfectly polarized electrons
without spin-flip, the spin current should be the same amount of the charge
current.

Equations (\ref{je(k)}) and (\ref{js(k)}) are the principal results of this
paper. One can obtain $d$-dimensional real space expressions by taking
inverse Fourier transform. This is one of the key advantages of our theory.
Our result is easily generalizable to $d$-dimensional result. As one shall
see, it turns out that the nonconservative part of $\vec{E}_s$ plays a
crucial role in a higher dimensional system.

First of all, we present 1D result. 1D real space expression of Eqs.
(\ref{je(k)}) and (\ref{js(k)}) are
\begin{eqnarray}
j_e(x,t)&=&0,\label{je(1D)}\\
j_s(x,t)&=&(1-P^2)\int dx' \frac{e^{-|x-x'|/\lambda_{sf}}}{2\lambda_{sf}} \sigma E_s(x',t).\label{js(1D)}
\end{eqnarray}
One can notice that the 1D charge current is perfectly canceled by diffusion
current for small spin-flip time limit. This is natural in the sense that,
for small spin-flip time, the system tends to be in equilibrium at each time
$t$. At equilibrium, the gradient of chemical potential vanishes, so does
charge current. However, spin current does not vanish by this reason because
of spin nonconserving process. Due to spin diffusion, it is natural that the
spin current becomes nonlocal with integration kernel width $\lambda_{sf}$.
Here, the factor $(1-P^2)$ seems unexpected. This factor comes from spin
diffusion effect, and should exist regardless of diffusion length scale. It
yields more cancelation for more polarized electrons. Eventually, for
$P=\pm1$, spin current also vanishes, which is actually required since
$\vec{j}_s=\vec{j}_e$ in the limit $P=\pm1$.

Equation (\ref{js(1D)}) behaves quite differently for two limiting cases. Let
$\lambda$ be the characteristic length scale (such as DW width) of localized
$E_s$. If $\lambda_{sf}\ll\lambda$,
$e^{|x-x'|/\lambda_{sf}}\approx2\lambda_{sf}\delta(x-x')$. Then,
\begin{equation}
j_s(x,t)\approx (1-P^2)\sigma E_s(x,t),
\end{equation}
which is local. Very short diffusion length cannot make the spin current
nonlocal. It is very interesting that $(1-P^2)$ factor does not disappear
even though diffusion effect is very small. The existence of diffusion makes
$(1-P^2)$ factor regardless of how the effect is strong or weak. For
$\lambda_{sf}\gg\lambda$,
\begin{equation}
j_s(x,t)\approx \sigma \overline{E}_s\frac{(1-P^2)\lambda}{2\lambda_{sf}}e^{-|x-X(t)|/\lambda_{sf}},
\end{equation}
where $\overline{E}_s$ is averaged SMF and $X(t)$ is the position of
localized $E_s$ (such as DW position). One can see that the current is highly
suppressed by the factor $\lambda/\lambda_{sf}$. In this highly diffusive
regime, spin current is also highly suppressed.

The main features of our result is similar to those of Ref.
\cite{Zhang10PRB}, except for vanishing charge current. They claim that
charge current can exist in general, while we find that Einstein's relation
prevents the existence of charge current in 1D.

\begin{figure}
\includegraphics[width=8.6cm]{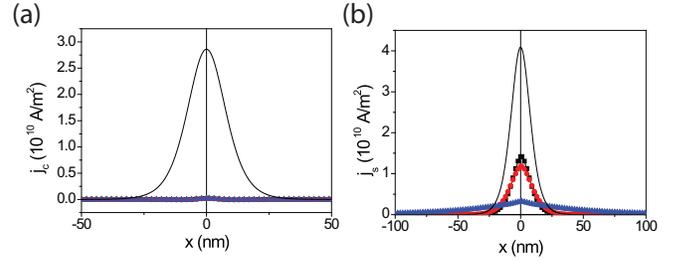}
\caption{
(color online) Results of micromagnetic simulation for (a) charge current and
(b) spin current induced by SMF generated from 1D DW oscillator,
which is fixed at $x=0$, but rotating with $\omega=10$GHz, in the presence of spin diffusion
with $\lambda_{sf}=0.5$nm (black rectangle), $5.0$nm (red circle), and $50$nm (blue triangle).
Solid line represents currents without diffusion. Here, DW width is set to be 10nm.
} \label{fig:simulation1D}
\end{figure}

We confirmed our analytic result by comparing it with micromagnetic
simulation for a 1D DW oscillator. The result is in Fig.
\ref{fig:simulation1D}. General features are the same as analytic result ; i)
charge and spin currents are highly suppressed by spin diffusion, ii) spin
current becomes nonlocal, iii) charge current almost vanish independently of
diffusion length, and iv) spin current is more suppressed by larger diffusion
length.

Now, we generalize the results to higher dimension. In 2D and 3D real spaces,
Eqs. (\ref{je(k)}) and (\ref{js(k)}) are converted to
\begin{subequations}
\label{je(3D)}
\begin{eqnarray}
\vec{j}_e^{2D}&=&P\sigma\vec{E}_s-\frac{P\sigma}{2\pi}\nabla\int d^2x'\ln\frac{|\vec{x}-\vec{x}'|}{\lambda_{sf}}\nabla'\cdot\vec{E}_s,\\
\vec{j}_e^{3D}&=&P\sigma\vec{E}_s+\frac{P\sigma}{4\pi}\nabla\int d^3x'\frac{\nabla'\cdot\vec{E}_s}{|\vec{x}-\vec{x}'|},
\end{eqnarray}
\end{subequations}
and
\begin{subequations}
\label{js(3D)}
\begin{eqnarray}
\vec{j}_s^{2D}&=&P\vec{j}_e^{2D}+(1-P^2)\sigma\vec{E}_s\nonumber\\
&&+\frac{(1-P^2)\sigma}{2\pi}\nabla\int d^2x'K_0\left(\frac{|\vec{x}-\vec{x}'|}{\lambda_{sf}}\right)\nabla'\cdot\vec{E}_s,\\
\vec{j}_s^{3D}&=&P\vec{j}_e^{3D}+(1-P^2)\sigma\vec{E}_s\nonumber\\
&&+\frac{(1-P^2)\sigma}{4\pi}\nabla\int d^3x'\frac{e^{-|\vec{x}-\vec{x}'|/\lambda_{sf}}}{|\vec{x}-\vec{x}'|}\nabla'\cdot\vec{E}_s,
\end{eqnarray}
\end{subequations}
respectively. Here, $K_0(x)$ is the zeroth order modified Bessel function of
the second kind. One can be uncomfortable because $\vec{j}_e^{2D}$ seems
dependent on $\lambda_{sf}$ while Eq. (\ref{je(k)}) is not. Here
$\lambda_{sf}$ is included because the argument of logarithmic function
should be dimensionless. In fact, one can easily check that
$\vec{j}_{e}^{2D}$ does not depend on $\lambda_{sf}$ by using $\ln
|\vec{x}-\vec{x}'|/\lambda_{sf}'=\ln
|\vec{x}-\vec{x}'|/\lambda_{sf}+$(constant) for any positive $\lambda_{sf}'$.

From now on, we present only 3D expressions but omit 2D, for simplicity. One
can easily obtain 2D expressions by replacing the integral kernels
$1/|\vec{x}-\vec{x}'|\rightarrow-2\ln|\vec{x}-\vec{x}'|/\lambda_{sf}$ and
$\exp(-|\vec{x}-\vec{x}'|/\lambda_{sf})/|\vec{x}-\vec{x}'|\rightarrow
2K_0\left(\frac{|\vec{x}-\vec{x}'|}{\lambda_{sf}}\right)$. The reason why the
expressions depend on dimension is nothing but the fact that the inverse
Fourier transforms, which give integration kernels, depend on dimension.
Hence, essential physics are the same for 2D and 3D except the mathematical
expressions of the integration kernels.

Overall features of the spin current in higher dimension is similar to 1D,
except for the existence of nonvanishing $P\vec{j}_e$ term. However, the main
feature of $\vec{j}_e$ is completely different from 1D case. First of all,
$\vec{j}_e$ does not vanish. We argued qualitatively why the charge current
vanishes in a 1D system by using chemical potential argument. However, in a
higher dimensional system, chemical potential cannot be defined in general
since $\vec{E}_s$ is nonconservative. Note that diffusion current
$D^\uda\nabla n^\uda$ is conservative. Note also that nonconservative field
cannot be canceled by conservative field. This is why the charge current
exists in a higher dimensional system. The nonlocal term in Eq.
(\ref{je(3D)}) can be interpreted as Coulomb potential under charge density
$-\nabla\cdot\vec{E}_s$. The canceled part of the charge current is nothing
but conservative Coulomb part of $\vec{E}_s$. Secondly, it is very
interesting that Eq. (\ref{je(3D)}) is converted after some algebra to
\begin{equation}
\vec{j}_e=\frac{P\sigma}{4\pi}\int d^3x'\frac{\nabla'\times \left(\nabla'\times\vec{E}_s\right)}{|\vec{x}-\vec{x}'|}.\label{je(3D)-curl}
\end{equation}
Now one can notice the importance of nonconservative part of $\vec{E}_s$
(nonvanishing $\nabla\times\vec{E}_s$) and the dependence of this on the
charge current. If $\nabla\times\vec{E}_s$ happens to be zero, the charge
current also vanishes, and this is consistent to the chemical potential
argument. Lastly, it is also interesting that charge current does not depend
on diffusion length. This is qualitatively understandable from the fact that
the effect of $\vec{E}_s$ is maximally canceled by diffusion current in small
spin-flip time regime, regardless of spin diffusion length. Here, maximal
cancelation is slightly different from perfect cancelation in 1D case in the
sense that the (conservative) diffusion current cannot cancel
$P\sigma\vec{E}_s$ perfectly in principle.

It might be ambiguous what the ``conservative part" of a vector field is
mathematically. Helmholtz's theorem guarantees that a spatially localized
vector field can be uniquely decomposed into conservative (curl-free) part
and solenoidal (divergence-free) part. Note that the second term in Eq.
(\ref{je(3D)}) is exactly the same as the formula of (negative of)
conservative part of the Helmholtz decomposition. Note also that the
resulting total current [Eq. (\ref{je(3D)-curl})] is divergence-free.
Therefore, the charge diffusion current and total charge current are
respectively given by conservative part and solenoidal part of Helmholtz
decomposition. One shall see in Sec. \ref{sec_side} that this claim is
generally valid for arbitrary boundary conditions.

\begin{figure}
\includegraphics[width=8.6cm]{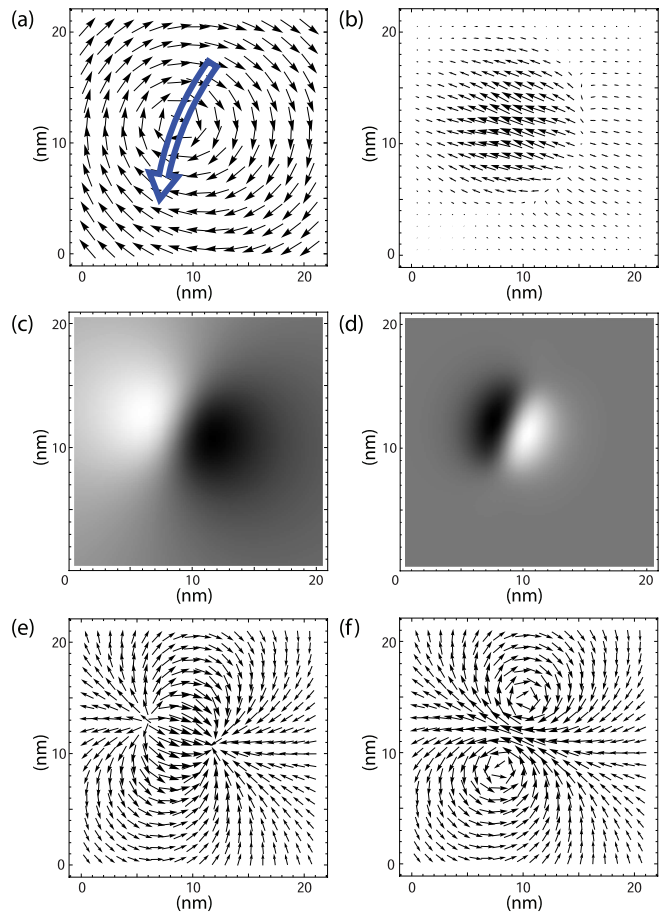}
\caption{
(color online) Results of the micromagnetic simulation for vortex resonant oscillation in
2D thin film. Here, the vortex core size is about 5nm, the resonant frequency is 605.5MHz, and $P=P_n=0.7$.
The size of arrows are log-scaled (size $\propto$ ln(1+norm)).
(a) Magnetization profile at the time when all calculation was performed.
The blue arrow denotes the direction of vortex core motion. (b) SMF. (c)
Accumulated charge density $-e\Delta n_e$ (divided by $\sigma\tau_{sf}/\lambda_{sf}^2$). The
maximum value (white) is $1.39\times 10^{-5}$V and the minimum value (black)
is $-1.62\times 10^{-5}$V. (d) $\nabla\cdot\vec{E}_s$. The maximum value
(white) is $8.43\times 10^{10}$V/m$^2$ and the minimum value (black) is
$-7.42\times 10^{10}$V/m$^2$. (e) Charge diffusion current, which is the
conservative part of SMF.
(f) Total charge current, which is the solenoidal part of SMF.
} \label{fig:simulation2D}
\end{figure}

To understand the maximal cancelation of the charge current qualitatively, it
would be very helpful to visualize the Helmholtz decomposition of the charge
current. We performed a micromagnetic simulation for vortex resonant
oscillation in 2D thin film. When the vortex DW wall [Fig.
\ref{fig:simulation2D}(a)] moves along the blue arrow, SMF is generated as
shown in Fig. \ref{fig:simulation2D}(b). The spatial dependence of SMF
induces charge accumulation as shown in Fig. \ref{fig:simulation2D}(c). Due
to the charge accumulation, charge diffusion current is generated as shown in
Fig. \ref{fig:simulation2D}(e).  Figure \ref{fig:simulation2D}(e) can be
qualitatively understood by Fig. \ref{fig:simulation2D}(d). Recall that the
diffusion current is given by Coulomb field generated by charge density
$-\nabla\cdot\vec{E}_s$. Thus, dipole-like nature of the charge density
$-\nabla\cdot\vec{E}_s$ [Fig. \ref{fig:simulation2D}(d)] implies dipole-like
field [Fig. \ref{fig:simulation2D}(e)]. Summing up Figs.
\ref{fig:simulation2D}(b) and \ref{fig:simulation2D}(e), one obtains total
current as Fig. \ref{fig:simulation2D}(f). Note that the total current is
definitely nonconservative (has finite curl). It is interesting that Fig.
\ref{fig:simulation2D}(f) is similar to magnetic field generated by two
separate conducting wires. This infers the solenoidal nature of the total
charge current. It is very interesting that total charge current behaves as
magnetic field rather than electric field.

\subsection{Magnetization dynamics and nonlocal damping tensor}

It is also important to see how the magnetization dynamics is changed by our
result. By analogue of Ref. \cite{Zhang04PRL}, it is obvious that the
modified LLG equation is described by Eq. (\ref{LLG}), in which Eq.
(\ref{js(3D)}) are added to $\vec{j}_s$ terms, rather than Eq. (\ref{js0}).
However, there exist two nontrivial features in the modified LLG equation.

The first one is spatial dependence of $\gamma$ and $\alpha$. It is important
to notice that $\gamma$ and $\alpha$ in Eq. (\ref{LLG}) are renormalized
parameters \cite{Zhang04PRL},
\begin{eqnarray}
\gamma&=&\gamma_0\left(1+\frac{n_s\mu_B}{M_s}\frac{1}{1+\beta^2}\right)^{-1},\\
\alpha&=&\frac{\gamma_0}{\gamma}\left(\alpha_0+\beta\frac{n_s\mu_B}{M_s}\frac{1}{1+\beta^2}\right),
\end{eqnarray}
where $\gamma_0$ and $\alpha_0$ are original parameters of the system. There
has not been any problem of this renormalization without spin accumulation,
but $n_s$ is no longer constant in the presence of spin accumulation. Hence,
$\gamma$ and $\alpha$ cannot be a simple constant in principle. At this
stage, it is more convenient to write down LLG equation without parameter
renormalization,
\begin{eqnarray}
&&\left(1+\frac{1}{1+\beta^2}\frac{n_s\mu_B}{M_s}\right)\frac{\partial\vec{M}}{\partial t}=-\gamma_0\vec{M}\times\vec{H}_{eff}\nonumber\\
&&+\left(1+\frac{\beta}{\alpha_0}\frac{1}{1+\beta^2}\frac{n_s\mu_B}{M_s}\right)\frac{\alpha_0}{M_s}\vec{M}\times\frac{\partial\vec{M}}{\partial t}\nonumber\\
&&+\frac{\mu_B}{eM_s}(\vec{j}_s\cdot\nabla)\vec{M}-\frac{\beta\mu_B}{eM_s^2}\vec{M}\times(\vec{j}_s\cdot\nabla)\vec{M}.\label{LLG(unrenormalized)}
\end{eqnarray}
In fact, the effect of $n_s$ variation is small. Note that
$n_s\mu_B/M_s=(n_0^\ua-n_0^\da)\mu_B/M_s\times\Delta n_s/(n_0^\ua-n_0^\da)$.
Here, the first factor is of the order of $10^{-2}$ and the second factor is
first order in SMF. Then, the effect should be very small compared to
ordinary first order effect of SMF. In addition, one shall see in Sec.
\ref{sec_dwmotion}, that symmetry can reduce the effect of $\Delta n_s$ in
collective coordinate level. In the case, the effect of $\Delta n_s$ vanishes
through odd function integration.

The next one is effective damping constant. Without applied electric field,
the modified LLG equation is obtained by taking Gilbert damping as a damping
tensor like Eq. (\ref{LLG(Z&Z)}). Then, it is interesting to see how the
damping tensor is generalized by spin diffusion effect. In the presence of
spin diffusion, local damping tensor $\mathcal{D}_{ij}$ becomes nonlocal
damping tensor $\mathcal{D}_{ij}(\vec{x},\vec{x}')$. By using Eq.
(\ref{js(3D)}), it turns out that the damping tensor in Eq. (\ref{LLG(Z&Z)})
is modified as
\begin{eqnarray}
\mathcal{D}_{ij}(x,x')&=&\alpha\delta_{ij}\delta(x-x')+\frac{(1-P^2)\eta}{M_s^4}\frac{e^{-|x-x'|/\lambda_{sf}}}{2\lambda_{sf}}\nonumber\\
&&\times(\vec{M}\times\partial_x\vec{M})_i(\vec{M}'\times\partial_x\vec{M}')_j,\label{damping tensor(1D)-main text}
\end{eqnarray}
for 1D and
\begin{eqnarray}
\mathcal{D}_{ij}(\vec{x},\vec{x}')&=&\alpha\delta_{ij}\delta(\vec{x}-\vec{x}')+\frac{\eta}{\sigma M_s^4}S_{kl}(\vec{x},\vec{x}')\nonumber\\
&&\times(\vec{M}\times\partial_k\vec{M})_i(\vec{M}'\times\partial_l\vec{M}')_j,\\
\mathcal{S}_{ij}(\vec{x},\vec{x}')&=&\sigma\delta_{ij}\delta(\vec{x}-\vec{x}')+\frac{P^2\sigma}{4\pi}\partial_i\partial_j\frac{1}{|\vec{x}-\vec{x}'|}\nonumber\\
&&+\frac{(1-P^2)\sigma}{4\pi}\partial_i\partial_j\frac{e^{-|\vec{x}-\vec{x}'|/\lambda_{sf}}}{|\vec{x}-\vec{x}'|}.
\end{eqnarray}
for 3D. Here, $\vec{M}'=\vec{M}(\vec{x}',t)$ and $\cdot$ is redefined as
inner product with respect to both coordinate basis and spatial basis,
\begin{eqnarray}
\left[\mathcal{D}\cdot \vec{f}(\vec{x})\right]_i=\sum_{j}\int d^dx'\mathcal{D}_{ij}(\vec{x},\vec{x}')f_j(\vec{x}'),\label{nonlocal tensor formula}
\end{eqnarray}
for a vector field $\vec{f}$. As a passing remark, the tensor $\mathcal{D}$
is indeed a \textit{damping} tensor in the sense that it decreases total
magnetic energy. It can be demonstrated by showing energy dissipation is
negative,
\begin{equation}
\frac{dE}{dt}=-\int \left(\vec{H}_{eff}\cdot\frac{\partial\vec{M}}{\partial t}\right)d^dx<0.\label{energy dissipation}
\end{equation}
The mathematical details related to Eqs. (\ref{damping tensor(1D)-main
text})-(\ref{energy dissipation}) are in Appendix \ref{sec_damping tensor}.

\section{Example : Domain wall motion \label{sec_dwmotion}}

In this section, we apply our theory to 1D DW motion. We find equation of
motion of collective coordinates $(X(t), \phi(t))$ of tail-to-tail transverse
DW. Here, $X(t)$ is DW position and $\phi(t)$ is tilting angle. Mathematical
details of obtaining the collective coordinate equation is described in
Appendix \ref{sec_dwmotion_appendix}.

Without SMF, DW motion is described by
\cite{Thiele73PRL,Jung08APL,Ryu09JAP,Malozemoff79,Schryer74JAP,Tatara06JPS,Tatara04PRL}
\begin{eqnarray}
-\frac{\partial\phi}{\partial t}+\frac{\alpha}{\lambda}\frac{\partial X}{\partial t}&=&-\frac{\beta b_J^0}{\lambda},\label{dw1 without SMF}\\
\frac{1}{\lambda}\frac{\partial X}{\partial t}+\alpha\frac{\partial \phi}{\partial t}&=&-\frac{b_J^0}{\lambda}-\frac{\gamma K_d}{M_s}\sin2\phi,\label{dw2 without SMF}
\end{eqnarray}
where $b_J^0=Pj\mu_B/eM_s$ for applied current $j$, $K_d$ represents dipole
field integration, and $\lambda$ is DW width. In the presence of SMF, Eqs.
(\ref{dw1 without SMF}) and (\ref{dw2 without SMF}) are modified as
\cite{Kim11CAP}
\begin{eqnarray}
-\frac{\partial\phi}{\partial t}+\frac{\alpha}{\lambda}\frac{\partial X}{\partial t}&=&-\frac{\beta b_J^0}{\lambda}-\frac{2\beta\eta}{3\lambda^2}\frac{\partial\phi}{\partial t},\label{dw1 with SMF}\\
\frac{1}{\lambda}\frac{\partial X}{\partial t}+\alpha\frac{\partial \phi}{\partial t}&=&-\frac{b_J^0}{\lambda}-\frac{2\eta}{3\lambda^2}\frac{\partial\phi}{\partial t}-\frac{\gamma K_d}{M_s}\sin2\phi.\label{dw2 with SMF}
\end{eqnarray}
One can expect that the effect of $\eta$ will be suppressed by spin
diffusion, so a renormalized parameter $\tilde{\eta}$ will replace $\eta$.
Thus, the expected equations of motion are
\begin{eqnarray}
-\frac{\partial\phi}{\partial t}+\frac{\alpha}{\lambda}\frac{\partial X}{\partial t}&=&-\frac{\beta b_J^0}{\lambda}-\frac{2\beta\tilde{\eta}}{3\lambda^2}\frac{\partial\phi}{\partial t},\label{dw1 with diffusion}\\
\frac{1}{\lambda}\frac{\partial X}{\partial t}+\alpha\frac{\partial \phi}{\partial t}&=&-\frac{b_J^0}{\lambda}-\frac{2\tilde{\eta}}{3\lambda^2}\frac{\partial\phi}{\partial t}-\frac{\gamma K_d}{M_s}\sin2\phi.\label{dw2 with diffusion}
\end{eqnarray}
As described in Appendix \ref{sec_dwmotion_appendix}, it turns out that Eqs.
(\ref{dw1 with diffusion}) and (\ref{dw2 with diffusion}) are indeed valid,
and the renormalized SMF parameter is given by
\begin{eqnarray}
\tilde{\eta}&=&(1-P^2)F(\zeta)\eta,\\
F(\zeta)&=&-\frac{3}{2}\zeta-\frac{3}{2}\zeta^2+\frac{3}{4}\zeta^3\left.\left(\frac{\Gamma'(z)}{\Gamma(z)}\right)\right|_{z=\frac{\zeta}{2}},
\end{eqnarray}
where $\zeta=\lambda/\lambda_{sf}$ and $\Gamma(z)$ is the Gamma function.

The renormalization is largely dependent on relative magnitude of DW width
and spin diffusion length. The asymptotic behavior of the function $F$ is
given by
\begin{equation}
F(\zeta)=\left\{\begin{array}{cc}
                  \frac{3}{2}\zeta-\frac{3}{2}\zeta^2+\frac{\pi^2\zeta^3}{8}+\mathcal{O}(\zeta^4) & \zeta\ll1, \\
                  1-\frac{4}{5\zeta^2}+\mathcal{O}(\zeta^{-4}) & \zeta\gg1.
                \end{array}
\right.
\end{equation}
For $\lambda\ll\lambda_{sf}$, Eqs. (\ref{dw1 with SMF}) and (\ref{dw2 with
SMF}) are reproduced except $(1-P^2)$ factor which should exist as already
discussed. For $\lambda\gg\lambda_{sf}$, the effect is highly suppressed by
spin diffusion, so the overall effect of SMF goes as
$\frac{1}{\lambda\lambda_{sf}}$ rather than $\frac{1}{\lambda^2}$.

\section{Further generalizations\label{sec_side}}

\subsection{Extended electric field - general boundary condition}

In the case that the magnetization dynamics is generated by applied electric
field, the electric field is no longer localized. Hence, it is necessary to
generalize our theory to non-localized electric field ; $\vec{E}_s$ does not
vanish at $|\vec{x}|\rightarrow\infty$. In this case, $\tilde{\mathcal{E}}$
does not give well-defined Fourier transform, but includes delta function
parts. Furthermore, in order to obtain Eq. (\ref{je(3D)-curl}), one obtains
additional boundary terms when integrating Eq. (\ref{je(3D)}) by parts.
Hence, the charge current may not be canceled by the diffusion current even
if the electric field is conservative.

The problem can be treated by Green's function method with given boundary
condition, as described in the last part of this section. However, it is hard
to catch the physical meaning, so we present more intuitive analysis for this
case. It is very convenient to use the linearity of our theory. Suppose
$\vec{E}^\uda$ can be decomposed into two components
$\vec{E}^\uda=\vec{E}_1^\uda+\vec{E}_2^\uda$. Then, since Eqs. (\ref{sdde2})
and (\ref{JJ}) are linear, the current $\vec{j}^\uda$ can also be decomposed
into two components $\vec{j}^\uda=\vec{j}_1^\uda+\vec{j}_2^\uda$, where
$\vec{j}_i^\uda$ is the current generated by $\vec{E}_i$.

To take advantage of this linearity, we decompose $\vec{E}$ into three
components as follows; any vector field $\vec{E}$ can be uniquely decomposed
into irrotational part $\vec{E}_1$, solenoidal part $\vec{E}_2$, and boundary
part $\vec{E}_3$,
\begin{eqnarray}
&&\vec{E}^\uda=\vec{E}_1^\uda+\vec{E}_2^\uda+\vec{E}_3^\uda,\\
&&\nabla\cdot\vec{E}_1^\uda\ne0,~\nabla\times\vec{E}_1^\uda=0,~\vec{n}\cdot\vec{E}_1^\uda|_{\vec{x}\in\partial}=0,\\
&&\nabla\cdot\vec{E}_2^\uda=0,~\nabla\times\vec{E}_2^\uda\ne0,~\vec{n}\cdot\vec{E}_2^\uda|_{\vec{x}\in\partial}=0,\\
&&\nabla\cdot\vec{E}_3^\uda=0,~\nabla\times\vec{E}_3^\uda=0,~\vec{n}\cdot\vec{E}_3^\uda|_{\vec{x}\in\partial}\ne0.
\end{eqnarray}
Here, $\partial$ means boundary and $\vec{n}$ denotes a normal unit vector
perpendicular to the boundary. From given $\vec{E}^\uda$, one can obtain
$\vec{E}_3^\uda$ by solving Laplace equation with Neumann boundary condition
and $\vec{E}_{1,2}^\uda$ by Helmholtz's theorem. As previously mentioned, our
result [Eq. (\ref{je(3D)-curl})] indicates that the conservative source part
$\vec{E}_1^\uda$ cannot contribute to the charge current while the
nonconservative solenoidal part $\vec{E}_2^\uda$ can give rise to a
nonvanishing contribution $\sigma^\ua\vec{E}_2^\ua+\sigma^\da\vec{E}_2^\da$.
[Eq. (\ref{je(3D)}) or Eq. (\ref{je(general)}) for general boundary
condition]

In the presence of nonvanishing boundary condition, it is important to
investigate the effect of $\vec{E}_3^\uda$ to the spin accumulation.
Fortunately, the source term of Eq. (\ref{sdde2}) depends on the divergence
of $\vec{E}_s^\uda$ only. Hence, $\vec{E}_3^\uda$ cannot contribute to the
spin accumulation since $\nabla\cdot\vec{E}_3^\uda=0$. Therefore,
$\vec{E}_3^\uda$ can only contribute to the currents via the first term in
Eq. (\ref{JJ}). Consequently, the expressions of charge and spin current
should be $\sigma^\ua \vec{E}_3^\ua\pm\sigma^\da \vec{E}_3^\da$ added to Eqs.
(\ref{je(3D)}) and (\ref{js(3D)}) (or Eqs. (\ref{je(1D)}) and
(\ref{js(1D)})).

It can be a good example that a constant spin independent electric field
$\vec{E}_{app}=E_{app}\hat{x}$ is applied in a 1D wire. In this case,
$\vec{E}_3^\uda=E_{app}\hat{x}$. Hence, the charge and spin current should be
$\sigma^\ua E_{app}\pm\sigma^\da E_{app}$ added to Eqs. (\ref{je(1D)}) and
(\ref{js(1D)}), so
\begin{eqnarray}
j_e&=&\sigma E_{app},\\
j_s&=&P\sigma E_{app}+(1-P^2)\int dx' \frac{e^{-|x-x'|/\lambda_{sf}}}{2\lambda_{sf}} \sigma E_s(x',t).
\end{eqnarray}
In this case, the charge current is not canceled by diffusion current, but
only charge current \textit{generated by SMF} (which is spatially localized)
is canceled.

\subsection{Charge neutrality and other spin independent potentials}

There is another important physical consequence of spin accumulation we have
ignored. This is Coulomb potential of accumulated electrons. Since Coulomb
potential is in general strong, electron tends to make local charge neutral.
Hence, giving a charge neutrality constraint,
\begin{equation}
\Delta n^\ua+\Delta n^\da=0 \label{charge neutrality}
\end{equation}
is a good approximation. To do this, it is convenient to introduce charge
neutrality potential $V_c$ and replace
$\vec{E}_s^\uda\rightarrow\vec{E}_s^\uda-\nabla V_c$ as suggested in Ref.
\cite{Tserkovnyak08PRB}. Then, the solution of the spin drift-diffusion
equation [Eq. (\ref{solution N2})] is modified by
\begin{equation}
\left(
  \begin{array}{c}
    \Delta \tilde{n}^\ua \\
    \Delta \tilde{n}^\da \\
  \end{array}
\right)=\frac{\tau_{sf}}{e}\Omega^{-1}\left(
                                        \begin{array}{cc}
                                          \sigma^\ua & 0 \\
                                          0 & \sigma^\da \\
                                        \end{array}
                                      \right)i\vec{k}\cdot\left(
                                                            \begin{array}{c}
                                                              \tilde{\vec{E}}_s-i\vec{k}\tilde{V}_c \\
                                                              -\tilde{\vec{E}}_s-i\vec{k}\tilde{V}_c \\
                                                            \end{array}
                                                          \right),
\end{equation}
where $V_c$ is found self-consistently to satisfy Eq. (\ref{charge
neutrality}). After straightforward algebra, one obtains charge neutrality
potential in terms of $\vec{E}_s$,
\begin{eqnarray}
\tilde{V}_c(\vec{k},t)&=&-\frac{Pi\vec{k}\cdot\tilde{\vec{E}}_s}{k^2}-\frac{(P_n-P)i\vec{k}\cdot\tilde{\vec{E}}_s}{k^2+\lambda_{sf}^{-2}},\\
V_c(\vec{x},t)&=&-\frac{P}{4\pi}\int d^3x'\frac{1}{|\vec{x}-\vec{x}'|}\nabla\cdot\vec{E}_s\nonumber\\
&&-\frac{P_n-P}{4\pi}\int\frac{e^{-|\vec{x}-\vec{x}'|/\lambda_{sf}}}{|\vec{x}-\vec{x}'|}\nabla\cdot\vec{E}_s.\label{V_c}
\end{eqnarray}

The effect of this potential is twofold. Firstly, $V_c$ gives additional
electrical current $-\sigma^\uda\nabla V_c$. In $k$-space,
\begin{eqnarray}
\Delta\tilde{\vec{j}}_c^\ua(\vec{k},t) &=&-\sigma^\ua i\vec{k}V_c,\label{dj1 by V_c}\\
\Delta\tilde{\vec{j}}_c^\da(\vec{k},t) &=&-\sigma^\da i\vec{k}V_c.\label{dj2 by V_c}
\end{eqnarray}
Secondly, $V_c$ affects diffusion current through $\nabla n^\uda$. After some
algebra, one obtains
\begin{eqnarray}
\left(
  \begin{array}{c}
    \Delta\tilde{\vec{j}}_D^\ua \\
    \Delta\tilde{\vec{j}}_D^\da \\
  \end{array}
\right)&=&i\vec{k}\left(
                   \begin{array}{cc}
                     \lambda_{sf}^{\ua2} & 0 \\
                     0 & \lambda_{sf}^{\da2} \\
                   \end{array}
                 \right)\Omega^{-1}\left(
                   \begin{array}{cc}
                     \sigma^\ua & 0 \\
                     0 & \sigma^\da \\
                   \end{array}
                 \right)\left(
                 \begin{array}{c}
                 k^2V_c \\
                 k^2V_c \\
                 \end{array}
                 \right)\nonumber\\
    &=&i\vec{k}V_c\left(
                    \begin{array}{c}
                      \sigma^\ua \\
                      \sigma^\da \\
                    \end{array}
                  \right),\label{djd by V_c}
\end{eqnarray}
which exactly cancels Eqs. (\ref{dj1 by V_c}) and (\ref{dj2 by V_c}). Hence,
there are no additional charge and spin currents attributed to $V_s$.

It is very interesting that one does not need to assume any specific form of
$V_c$ to reach Eq. (\ref{djd by V_c}). The mathematical origin of the exact
cancelation is that the additional force is conservative (described by
$i\vec{k}\times$(scalar)) and spin independent. A spin independent potential
generates additional current, but spin accumulation adjusts fast to make
diffusion current cancel it. Of course, the modified spin accumulation
$n^\uda$ affects LLG equation by Eq. (\ref{LLG(unrenormalized)}), but we
claimed that this is ignorable. Consequently, any spin independent potential
cannot modify the main features of our result. As another side remark, this
cancelation can be verified for exact Coulomb potential even without using
the charge neutrality approximation Eqs. (\ref{charge
neutrality})-(\ref{V_c}).

\subsection{Arbitrary boundary}

Since SMF is usually strongly localized, we have considered an infinite
boundary problem. However, for a finite system, the expression of the charge
and spin currents should be slightly modified. Note that the key part of our
theory is to find real space expression of $\Omega^{-1}$. (See Eq.
(\ref{solution N2})). In real space, $\Omega$ is a differential operator,
\begin{equation}
\Omega_{real}=\left(
         \begin{array}{cc}
           -\lambda_{sf}^{\ua2}\nabla^2+\frac{n_0^\da}{n_0^\ua+n_0^\da} & -\frac{n_0^\ua}{n_0^\ua+n_0^\da} \\
           -\frac{n_0^\da}{n_0^\ua+n_0^\da} & -\lambda_{sf}^{\da2}\nabla^2+\frac{n_0^\ua}{n_0^\ua+n_0^\da} \\
         \end{array}
       \right).
\end{equation}
Hence, the problem is to find the inverse operator, i.e., the Green's
function corresponding $\Omega_{real}$ at the given geometry.

In order to find the Green's function of $\Omega_{real}$, one can get a hint
from Eq. (\ref{omega^-1}). Firstly, let $G_L(\vec{x},\vec{x}')$ and
$G_H(\vec{x},\vec{x}')$ be respectively the Green's function corresponding
Laplacian $\nabla^2$ and modified Helmholtz operator
$\nabla^2-\lambda_{sf}^{-2}$ for the given geometry. In order to obtain the
Green's function, it is plausible to replace
$1/k^2\rightarrow-G_L(\vec{x},\vec{x}')$ and
$1/(k^2+\lambda_{sf}^{-2})\rightarrow-G_H(\vec{x},\vec{x}')$ in Eq.
(\ref{omega^-1}). Then, one obtains
\begin{equation}
G_{\Omega}(\vec{x},\vec{x}')=-\frac{1}{\sigma}\left(
                     \begin{array}{cc}
                       \frac{\sigma^\ua G_L+\sigma^\da G_H}{\lambda_{sf}^{\ua2}} & \frac{\sigma^\ua G_L-\sigma^\ua G_H}{\lambda_{sf}^{\ua2}} \\
                       \frac{\sigma^\da G_L-\sigma^\da G_H}{\lambda_{sf}^{\da2}} & \frac{\sigma^\da G_L+\sigma^\ua G_H}{\lambda_{sf}^{\da2}} \\
                     \end{array}
                   \right).
\end{equation}
One can verify this is indeed the Green's function of $\Omega_{real}$ by
showing $\Omega_{real}G_{\Omega}=\delta(\vec{x}-\vec{x}')$. Then, one
straightforwardly concludes that the generalized expressions of charge and
spin current [Eqs. (\ref{je(3D)})-(\ref{js(3D)})] are given by
\begin{eqnarray}
\vec{j}_e&=&P\sigma\vec{E}_s-P\sigma\nabla\int d^dx'G_L(\vec{x},\vec{x}')\nabla'\cdot\vec{E}_s,\label{je(general)}\\
\vec{j}_s&=&P\vec{j}_e+(1-P^2)\sigma\vec{E}_s\nonumber\\
&&-(1-P^2)\sigma\nabla\int d^dx' G_H(\vec{x},\vec{x}')\nabla'\cdot\vec{E}_s.
\end{eqnarray}
For infinite boundary with vanishing boundary condition, $4\pi
G_L=-1/|\vec{x}-\vec{x}'|$ and $4\pi
G_H=-e^{-|\vec{x}-\vec{x}'|/\lambda_{sf}}/|\vec{x}-\vec{x}'|$, so Eqs.
(\ref{je(3D)})-(\ref{js(3D)}) are reproduced.

\section{Conclusion\label{sec_conclusion}}

By constructing spin drift-diffusion equation from the equation of motion of
conduction electrons, we studied the effect of SMF in the presence of spin
accumulation, spin diffusion and spin flip scattering, which were ignored
\cite{Zhang09PRL} or considered only in 1D \cite{Zhang10PRB} in previous
theories. It turns out that, in realistic regime, the conservative part of
charge current is canceled by diffusion current, and spin current becomes
nonlocal. Consequently, the magnetization dynamics is affected by
\textit{nonlocal} Gilbert damping tensor, instead of the previously reported
(local) Gilbert damping tensor. By calculating spin-transfer torque generated
by nonlocal spin current, we obtained the explicit expressions of the
nonlocal Gilbert damping tensor.

Different from the previous work focusing on 1D \cite{Zhang10PRB}, we also
obtained the results for 2D and 3D as well as 1D. In a 1D system, the results
of the previous theory are reproduced. It turns out that Einstein's relation
prevents the existence of charge current, but for parameter sets which do not
satisfy the Einstein's relation, the charge current can be induced by the
spin motive force. After generalizing the results to higher dimension, we
find that the nonconservative part of SMF plays an important role in charge
and spin currents.

As an illustration of suppression of the effect of SMF, we demonstrated
equations of motion of collective coordinate of 1D current-induced DW motion.
We found that spin diffusion renormalizes SMF depending on the relative
magnitude of DW width and spin diffusion length.

We also investigated the system under spatially extended (non-localized)
electric field. In this case, it turns out that the spatially extended part
of the electric field can contribute to the charge current even though it is
conservative. However, the existence of spatially extended part of the
electric field cannot alter the result that the charge current
\textit{generated by (localized) SMF} cannot include conservative part.

Our result is solid in the sense that our principal results are not changed
by any spin independent potential. A spin independent potential modifies
current via additional electric field, but spin density rapidly adjusts to
make diffusion current cancel it. Consequently, a spin independent potential
can modify spin density, but not current.

\begin{acknowledgements}
This work is financially supported by the NRF (2009-0084542, 2010-0014109,
2010-0023798), KRF (KRF-2009-013-C00019) and BK21. KWK acknowledges the
financial support by TJ Park.
\end{acknowledgements}

\begin{appendix}
\section{Absence of Time Variation of Equilibrium Number Density\label{sec_appen_n0}}
In this section, we show that $e^{-\Omega
t/\tau_{sf}}\tilde{\mathcal{N}}(\vec{k},0)$ does not have time dependence,
where $\Omega$ is given by Eq. (\ref{omega}) and
$\tilde{\mathcal{N}}(\vec{k},0)=(2\pi)^{d/2}\delta(\vec{k})(n_0^\ua~n_0^\da)^T$.
It suffices to calculate $e^{-\Omega t/\tau_{sf}}$ at $k=0$ because of
$\delta(\vec{k})$ factor. It is easy to show that $\Omega$ is idempotent for
$k=0$. In other words, $\Omega^n|_{k=0}=\Omega|_{k=0}$. Then, for $k=0$,
\begin{eqnarray}
e^{-\Omega t/\tau_{sf}}&=&\sum_{n=0}^\infty \frac{\Omega^n}{n!}\left(-\frac{t}{\tau_{sf}}\right)^n
=I+\sum_{n=1}^\infty \frac{\Omega}{n!}\left(-\frac{t}{\tau_{sf}}\right)^n,\nonumber\\
&=&I+\Omega(e^{-t/\tau_{sf}}-1).\label{A. exponent}
\end{eqnarray}
where $I$ is the identity matrix. Since
\begin{equation}
\Omega\tilde{\mathcal{N}}(\vec{k},0)=(2\pi)^{d/2}\delta(\vec{k})\Omega\left(
                                                                          \begin{array}{c}
                                                                            n_0^\ua \\
                                                                            n_0^\da \\
                                                                          \end{array}
                                                                        \right)=0
%&=&(2\pi)^{d/2}\delta(\vec{k})\left(
%                                \begin{array}{cc}
%                                  \frac{n_0^\da}{n_0^\ua+n_0^\da} & -\frac{n_0^\ua}{n_0^\ua+n_0^\da} \\
%                                  -\frac{n_0^\da}{n_0^\ua+n_0^\da} & \frac{n_0^\ua}{n_0^\ua+n_0^\da} \\
%                                \end{array}
%                              \right)\left(
%                              \begin{array}{c}
%                              n_0^\ua \\
%                              n_0^\da \\
%                              \end{array}\right)\nonumber\\&=&0,
\end{equation}
the second term of Eq. (\ref{A. exponent}) vanishes after applied to
$\tilde{\mathcal{N}}(\vec{k},0)$. Finally, one obtains $e^{-\Omega
t/\tau_{sf}}\tilde{\mathcal{N}}(\vec{k},0)=\tilde{\mathcal{N}}(\vec{k},0)$.

Note that the result is obtained without the approximation that $\tau_{sf}$
is small.

\section{Derivation of Eq. (\ref{delta})\label{sec_appen_delta}}

The idea is based on the delta sequence
\begin{equation}
\lim_{n\rightarrow\infty}ne^{-nt}\Theta(t)=\delta(t),
\end{equation}
for natural number $n$. This can be generalized as
\begin{equation}
\lim_{u\rightarrow\infty}ne^{-uzt}\Theta(t)=\delta(t),
\end{equation}
where $z$ is a complex number satisfying $\Re[z]>0$. This generalization is
obvious in that $uze^{-uzt}$ is localized at $t=0$ and $\int_{-\infty}^\infty
uze^{-uzt}\Theta(t)dt=1$.

We now generalize this relation to matrices. For a diagonalizable matrix $M$
with eigenvalues $\lambda_i$ satisfying $\Re[\lambda_{i}]>0$, we claim that
\begin{equation}
\lim_{u\rightarrow\infty}uMe^{-uMt}\Theta(t)=\delta(t)I.
\end{equation}
The proof is simple. Since $M$ is diagonalizable, one can write
\begin{equation}
uM=Q\left(%
\begin{array}{ccc}
  u\lambda_1 & 0 & \cdots \\
  0 & u\lambda_2 & \cdots \\
  \vdots & \vdots & \ddots \\
\end{array}%
\right)Q^{-1},
\end{equation}
for some $Q$. Then
\begin{eqnarray}
&&uMe^{-uMx}\Theta(x)\nonumber\\
&&=Q\left(%
\begin{array}{ccc}
  u\lambda_1e^{-u\lambda_1}\Theta(x) & 0 & \cdots \\
  0 & u\lambda_2e^{-u\lambda_2}\Theta(x) & \cdots \\
  \vdots & \vdots & \ddots \\
\end{array}%
\right)Q^{-1}\nonumber\\
&&\rightarrow Q\left(%
\begin{array}{ccc}
  \delta(x) & 0 & \cdots \\
  0 & \delta(x) & \cdots \\
  \vdots & \vdots & \ddots \\
\end{array}%
\right)Q^{-1}=\delta(x)I,
\end{eqnarray}
as $u\rightarrow\infty$.

By taking $M=\Omega$ and $u=1/\tau_{sf}$,
\begin{equation}
\lim_{\tau_{sf}\rightarrow +0}\frac{\Omega}{\tau_{sf}}e^{-\Omega t/\tau_{sf}}\Theta(t)=\delta(t)I,
\end{equation}
so, for small $\tau_{sf}$,
\begin{equation}
\frac{\Omega}{\tau_{sf}}e^{-\Omega t/\tau_{sf}}\Theta(t)\approx\delta(t)I,
\end{equation}
which is exactly Eq. (\ref{delta}).

As a passing remark, one should check that the real parts of eigenvalues of
$\Omega$ are really positive. It is guaranteed by $\mathrm{Tr}~\Omega>0$ and
$\det\Omega>0$ for any nonzero $\vec{k}$. However, the approximation Eq.
(\ref{delta}) makes singularity at $\vec{k}=0$. Fortunately, this singularity
is removed by $i\vec{k}$ factor in Eq. (\ref{solution N}).

\section{Nonlocal damping tensor and energy dissipation\label{sec_damping tensor}}

\subsection{1D damping tensor}
Starting from
\begin{eqnarray}
j_s&=&(1-P^2)\int dx' \frac{e^{-|x-x'|/\lambda_{sf}}}{2\lambda_{sf}}\sigma E_s(x',t),\\
E_s&=&\frac{\hbar}{2eM_s^3}\partial_t\vec{M}\cdot(\partial_x\vec{M}\times\vec{M}),
\end{eqnarray}
spin-transfer torque driven by SMF is given by
\begin{eqnarray}
\frac{\mu_B}{eM_s}j_s\partial_x\vec{M}&=&(1-P^2)\frac{\eta}{M_s^3}\partial_x\vec{M}\nonumber\\
&&\times\int dx' \frac{e^{-|x-x'|/\lambda_{sf}}}{2\lambda_{sf}}\partial_t\vec{M}'\cdot(\partial_x\vec{M}'\times\vec{M}')\nonumber\\
%&=&(1-P^2)\frac{\eta}{M_s^5}\vec{M}\times(\vec{M}\times\partial_x\vec{M})\nonumber\\
%&&\times\int dx' \frac{e^{-|x-x'|/\lambda_{sf}}}{2\lambda_{sf}}(\vec{M}'\times\partial_x\vec{M}')\cdot\partial_t\vec{M}'\nonumber\\
&=&\frac{\vec{M}}{M_s}\times\int dx'\left[(1-P^2)\frac{\eta}{M_s^4}\frac{e^{-|x-x'|/\lambda_{sf}}}{2\lambda_{sf}}\right.\nonumber\\
&&\left.(\vec{M}\times\partial_x\vec{M})(\vec{M}'\times\partial_x\vec{M}')\right]\cdot\partial_t\vec{M}'.
\end{eqnarray}
Then, comparing with Eqs. (\ref{LLG(Z&Z)}) and (\ref{nonlocal tensor
formula}),
\begin{eqnarray}
\mathcal{D}_{ij}(x,x')&=&\alpha\delta_{ij}\delta(x-x')+\frac{(1-P^2)\eta}{M_s^4}\frac{e^{-|x-x'|/\lambda_{sf}}}{2\lambda_{sf}}\nonumber\\
&&\times(\vec{M}\times\partial_x\vec{M})_i(\vec{M}'\times\partial_x\vec{M}')_j.\label{damping tensor(1D)}
\end{eqnarray}
It is interesting to note that
\begin{eqnarray}
\lim_{\lambda_{sf}\rightarrow0}\mathcal{D}_{ij}(x-x')&=&\delta(x-x')\left[\alpha\delta_{ij}+\frac{(1-P^2)\eta}{M_s^4}\right.\nonumber\\
&&\left.\times(\vec{M}\times\partial_x\vec{M})_i(\vec{M}\times\partial_x\vec{M})_j\right].
\end{eqnarray}
This is exactly the previous result Eq. (\ref{damping(Z&Z)}) except $(1-P^2)$
factor, which should exist regardless of diffusion strength.

Now, the remaining step is to calculate energy dissipation. Energy
dissipation is calculated by the integration energy density dissipation
$-\vec{H}_{eff}\cdot\partial_t\vec{M}$. From now on, the subscription
$_{eff}$ is dropped until this section ends. Energy dissipation by a nonlocal
damping tensor $\mathcal{D}$ is given by
\begin{eqnarray}
\frac{dE}{dt}&=&-\int dx \vec{H}\cdot\left[\frac{\vec{M}}{M_s}\times\mathcal{D}\cdot\partial_t\vec{M}\right]\nonumber\\
&=&\frac{1}{M_s}\int dx (\vec{M}\times\vec{H})\cdot\mathcal{D}\cdot\partial_t\vec{M}\nonumber\\
&\approx&-\frac{\gamma}{M_s}\int dx (\vec{M}\times\vec{H})\cdot\mathcal{D}\cdot(\vec{M}\times\vec{H})\nonumber\\
&=&-\frac{\gamma}{M_s}\iint dxdx' (\vec{M}\times\vec{H})_i\mathcal{D}_{ij}(x,x')(\vec{M}'\times\vec{H}')_j,
\end{eqnarray}
up to first order. Here, $\vec{H}'=\vec{H}(\vec{x}',t)$. The first term in
Eq. (\ref{damping tensor(1D)}) gives definitely negative $dE/dt$. The second
term, which comes from SMF, gives
\begin{eqnarray}
\frac{dE_{SMF}}{dt}&=&-\frac{(1-P^2)\eta\gamma}{M_s}\iint dxdx'(\vec{H}\cdot\partial_x\vec{M})\nonumber\\
&&\times\frac{e^{-|x-x'|/\lambda_{sf}}}{2\lambda_{sf}}(\vec{H}'\cdot\partial_x\vec{M}').
\end{eqnarray}
In order to show $dE/dt<0$, it is sufficient to show that $\iint dxdx'
f(x)e^{-a|x-x'|}f(x')$ is positive for real function $f$. This is verified by
Parseval's relation and convolution theorem of Fourier transform.
\begin{eqnarray}
&&\iint dxdx' f(x)e^{-a|x-x'|}f(x')\nonumber\\
&&=\int dx f(x)^* \left[\int dx' e^{-a|x-x'|}f(x')\right]\nonumber\\
&&=\int dk \mathcal{F}[f(x)](k)^*\mathcal{F}\left[\int dx' e^{-a|x-x'|}f(x')\right](k)\nonumber\\
&&=\sqrt{2\pi}\int dk \mathcal{F}[f(x)](k)^*\mathcal{F}[e^{-a|x|}](k)\mathcal{F}[f(x)](k)\nonumber\\
&&=2a\int dk \frac{\left|\mathcal{F}[f(x)](k)\right|^2}{k^2+a^2}>0.
\end{eqnarray}
This implies $dE/dt<0$.

\subsection{3D damping tensor}
From now on, Einstein's convention is used. Componentwise expressions of spin
current and SMF for a 3D system are
\begin{eqnarray}
j_{s,i}&=&\sigma E_{s,i}+\frac{P^2\sigma}{4\pi}\partial_i\int d^3x'\frac{\partial_j' E_{s,j}}{|\vec{x}-\vec{x}'|}\nonumber\\
&&+\frac{(1-P^2)\sigma}{4\pi}\partial_i\int d^3x'\frac{e^{-|\vec{x}-\vec{x}'|/\lambda_{sf}}}{|\vec{x}-\vec{x}'|}\partial_j' E_{s,j},\\
E_{s,i}&=&\frac{\hbar}{2eM_s^3}\partial_t\vec{M}\cdot(\partial_i\vec{M}\times\vec{M}).
\end{eqnarray}
By integrating by parts in order to remove derivatives in front of $E_{s,j}$,
\begin{eqnarray}
j_{s,i}&=&\sigma E_{s,i}-\frac{P^2\sigma}{4\pi}\partial_i\int d^3x'E_{s,j}(\vec{x}')\partial_j' \frac{1}{|\vec{x}-\vec{x}'|}\nonumber\\
&&-\frac{(1-P^2)\sigma}{4\pi}\partial_i\int d^3x'E_{s,j}(\vec{x}')\partial_j'\frac{e^{-|\vec{x}-\vec{x}'|/\lambda_{sf}}}{|\vec{x}-\vec{x}'|},\nonumber\\
&=&\sigma E_{s,i}+\frac{P^2\sigma}{4\pi}\int d^3x'E_{s,j}(\vec{x}')\partial_i'\partial_j' \frac{1}{|\vec{x}-\vec{x}'|}\nonumber\\
&&+\frac{(1-P^2)\sigma}{4\pi}\int d^3x'E_{s,j}(\vec{x}')\partial_i'\partial_j'\frac{e^{-|\vec{x}-\vec{x}'|/\lambda_{sf}}}{|\vec{x}-\vec{x}'|}.
\end{eqnarray}
Here, it is convenient to introduce nonlocal conductivity tensor
$\mathcal{S}$,
\begin{eqnarray}
j_{s,i}&=&\int d^3x' \mathcal{S}_{ij}(\vec{x},\vec{x}')E_{s,j}(\vec{x}'),\\
\mathcal{S}_{ij}(\vec{x},\vec{x}')&=&\sigma\delta_{ij}\delta(\vec{x}-\vec{x}')+\frac{P^2\sigma}{4\pi}\partial_i'\partial_j'\frac{1}{|\vec{x}-\vec{x}'|}\nonumber\\
&&+\frac{(1-P^2)\sigma}{4\pi}\partial_i'\partial_j'\frac{e^{-|\vec{x}-\vec{x}'|/\lambda_{sf}}}{|\vec{x}-\vec{x}'|}.
\end{eqnarray}
Now, spin-transfer torque driven by SMF is given by
\begin{eqnarray}
&&\frac{\mu_B}{eM_s}(\vec{j}_s\cdot\nabla)\vec{M}\nonumber\\
&&=\frac{\mu_B}{eM_s}\partial_i\vec{M}\int d^3x' \cdot\mathcal{S}_{ij}(\vec{x},\vec{x}')E_{s,j}(\vec{x}')\nonumber\\
%&&=\frac{\eta}{\sigma M_s^3}\partial_i\vec{M}\int d^3x' \cdot\mathcal{S}_{ij}(\vec{x},\vec{x}')(\partial_j\vec{M}'\times\vec{M}')\cdot\partial_t\vec{M}'\nonumber\\
&&=\frac{\vec{M}}{M_s}\times\left[\frac{\eta}{\sigma M_s^4}(\vec{M}\times\partial_i\vec{M})\right.\nonumber\\
&&\phantom{=}\left.\times\int d^3x' \cdot\mathcal{S}_{ij}(\vec{x},\vec{x}')(\vec{M}'\times\partial_j\vec{M}')\cdot\partial_t\vec{M}'\right].
\end{eqnarray}
Hence, the nonlocal damping tensor for 3D is given by
\begin{eqnarray}
\mathcal{D}_{ij}(\vec{x},\vec{x}')&=&\alpha\delta_{ij}\delta(\vec{x}-\vec{x}')+\frac{\eta}{\sigma M_s^4}S_{kl}(\vec{x},\vec{x}')\nonumber\\
&&\times(\vec{M}\times\partial_k\vec{M})_i(\vec{M}'\times\partial_l\vec{M}')_j.
\end{eqnarray}

Now, we calculate energy dissipation. This is an analogue of the previous
section. After some algebra,
\begin{eqnarray}
\frac{dE_{SMF}}{dt}&=&-\frac{\gamma\eta}{\sigma M_s}\iint d^3xd^3x'\mathcal{S}_{kl}(\vec{x},\vec{x}')\nonumber\\
&&\times(\vec{H}\cdot\partial_k\vec{M})(\vec{H}'\cdot\partial_l\vec{M}').
\end{eqnarray}
In order to show $dE_{SMF}/dt<0$, one should verify
\begin{equation}
\iint d^3xd^3x'(\vec{H}\cdot\partial_i\vec{M})\mathcal{S}_{ij}(\vec{x},\vec{x}')(\vec{H}'\cdot\partial_j\vec{M}')>0.
\end{equation}
Note that $\mathcal{S}_{ij}(\vec{x},\vec{x}')$ is a function of
$(\vec{x}-\vec{x}')$. It is convenient to write
$\mathcal{S}_{ij}=\mathcal{S}_{ij}(\vec{x}-\vec{x}')$ at this stage. Similar
to the previous subsection, by using Parseval's theorem and convolution
theorem of Fourier transform,
\begin{eqnarray}
&&\iint d^3xd^3x'(\vec{H}\cdot\partial_i\vec{M})\mathcal{S}_{ij}(\vec{x}-\vec{x}')(\vec{H}'\cdot\partial_j\vec{M}')\nonumber\\
&&=\int d^3k\mathcal{F}[\vec{H}\cdot\partial_i\vec{M}](\vec{k})^*\tilde{\mathcal{S}}_{ij}(\vec{k})\mathcal{F}[\vec{H}\cdot\partial_j\vec{M}](\vec{k}).\label{C18}
\end{eqnarray}
Here, $\tilde{\mathcal{S}}_{ij}(\vec{k})$ is given by
\begin{equation}
\tilde{\mathcal{S}}_{ij}(\vec{k})=\frac{\sigma}{(2\pi)^{3/2}}\left[\delta_{ij}-P^2\frac{k_ik_j}{k^2}
-(1-P^2)\frac{k_ik_j}{k^2+\lambda_{sf}^{-2}}\right].
\end{equation}
Note that the integrand in Eq. (\ref{C18}) is an expectation value of
$3\times3$ matrix $\tilde{\mathcal{S}}_{ij}(\vec{k})$ with respect to vector
$\mathcal{F}[\vec{H}\cdot\partial_i\vec{M}](\vec{k})$. Since
$\tilde{\mathcal{S}}_{ij}(\vec{k})$ is a Hermitian matrix, the integrand is
always positive if the eigenvalues of $\tilde{\mathcal{S}}_{ij}(\vec{k})$ are
positive. Recall that the eigenvalues of matrix $k_ik_j$ is given by
$(0,0,k^2)$. The corresponding eigenvectors are definitely the eigenvectors
of $\tilde{\mathcal{S}}_{ij}(\vec{k})$, so the eigenvalues of
$\tilde{\mathcal{S}}_{ij}(\vec{k})$ is given by
$\left(\frac{\sigma}{(2\pi)^{3/2}}, \frac{\sigma}{(2\pi)^{3/2}},
\frac{\sigma}{(2\pi)^{3/2}}\frac{\lambda_{sf}^{-2}(1-P^2)}{k^2+\lambda_{sf}^{-2}}\right)$,
which are all positive. This proves $dE/dt<0$.

\section{1D DW motion in the presence of spin diffusion\label{sec_dwmotion_appendix}}
\subsection{Collective coordinate equation of 1D DW for space-time dependent spin current and spin density}
The starting equation is 1D version of Eq. (\ref{LLG(unrenormalized)}).
\begin{eqnarray}
&&\left(1+\frac{1}{1+\beta^2}\frac{n_s\mu_B}{M_s}\right)\frac{\partial\vec{M}}{\partial t}=-\gamma_0\vec{M}\times\vec{H}_{eff}\nonumber\\
&&+\left(1+\frac{\beta}{\alpha_0}\frac{1}{1+\beta^2}\frac{n_s\mu_B}{M_s}\right)\frac{\alpha_0}{M_s}\vec{M}\times\frac{\partial\vec{M}}{\partial t}\nonumber\\
&&+b_J\frac{\partial\vec{M}}{\partial x}-\beta b_J\frac{\vec{M}}{M_s}\times\frac{\partial\vec{M}}{\partial x},
\end{eqnarray}
where $b_J=\mu_Bj_s/eM_s$. The main difference from the conventional approach
is that $b_J$ and $n_s$ are space-time dependent. The equation is rewritten
as
\begin{equation}
\gamma_0\vec{M}\times\vec{H}_{tot}=0,\label{D2}
\end{equation}
where
\begin{eqnarray}
\gamma_0\vec{H}_{tot}&=&-\left(1+\frac{1}{1+\beta^2}\frac{n_s\mu_B}{M_s}\right)\frac{\vec{M}}{M_s^2}\times\frac{\partial\vec{M}}{\partial t}+\gamma_0\vec{H}_{eff}\nonumber\\
&&-\left(\alpha_0+\frac{\beta}{1+\beta^2}\frac{n_s\mu_B}{M_s}\right)\frac{1}{M_s}\frac{\partial\vec{M}}{\partial t}\nonumber\\
&&+b_J\frac{\vec{M}}{M_s^2}\times\frac{\partial\vec{M}}{\partial x}+\frac{\beta b_J}{M_s}\frac{\partial\vec{M}}{\partial x}.
\end{eqnarray}
We set the effective magnetic field as
$\vec{H}_{eff}=\frac{2A}{M_s^2}\frac{\partial^2\vec{M}}{\partial
x^2}+\frac{H_KM_x}{M_s}\hat{x}+\vec{H}_d$, where $A$ and $K_d$ represents
exchange coupling and anisotropy, and $\vec{H}_d$ is dipole field. In
addition, the magnetization profile is set to be tail-to-tail transverse
wall,
\begin{eqnarray}
\vec{M}&=&M_s(\cos\theta,\sin\theta\cos\phi,\sin\theta\sin\phi),\\
\sin\theta&=&\secha\left(\frac{x-X(t)}{\lambda}\right),\\
\cos\theta&=&\tanh\left(\frac{x-X(t)}{\lambda}\right),\\
\phi&=&\phi(t),
\end{eqnarray}
where $X(t)$ is DW position and $\phi(t)$ is tilting angle.

Note that Eq. (\ref{D2}) implies $\vec{H}_{tot}=a\vec{M}$ for some $a$. Then,
one can define force density
$f=-\vec{H}_{tot}\cdot\frac{\partial\vec{M}}{\partial X}$ and torque density
$\tau=-\vec{H}_{tot}\cdot\frac{\partial\vec{M}}{\partial \phi}$, which are
identically zero. Then, collective coordinate equation is given by
calculating total force and total torque,
\begin{eqnarray}
F&\equiv&\int fdx=0,\\
T&\equiv&\int \tau dx=0.
\end{eqnarray}
Each equation implies respectively,
\begin{eqnarray}
&&-\frac{\partial\phi}{\partial t}+\frac{\alpha}{\lambda}\frac{\partial X}{\partial t}+\frac{1}{1+\beta^2}\left(-\frac{\partial\phi}{\partial t}+\frac{\beta}{\lambda}\frac{\partial X}{\partial t}\right)\nonumber\\
&&\times\int\frac{\Delta n_s\mu_B}{M_s}\sech^2\left(\frac{x-X}{\lambda}\right)\frac{dx}{2\lambda}\nonumber\\
&&=-\frac{\beta b_J^0}{\lambda}-\frac{\beta}{\lambda}\int b_J^s \sech^2\left(\frac{x-X}{\lambda}\right)\frac{dx}{2\lambda},\label{D10}\\
&&\frac{1}{\lambda}\frac{\partial X}{\partial t}+\alpha\frac{\partial \phi}{\partial t}+\frac{1}{1+\beta^2}\left(\frac{1}{\lambda}\frac{\partial X}{\partial t}+\beta\frac{\partial \phi}{\partial t}\right)\nonumber\\
&&\times\int \frac{\Delta n_s\mu_B}{M_s}\sech^2\left(\frac{x-X}{\lambda}\right)\frac{dx}{2\lambda}\nonumber\\
&&=-\frac{b_J^0}{\lambda}-\int\frac{b_J^s}{\lambda}\sech^2\left(\frac{x-X}{\lambda}\right)\frac{dx}{2\lambda}-\frac{\gamma K_d}{M_s}\sin 2\phi.\label{D11}
\end{eqnarray}
where $b_J^0$ is space-time independent part of $b_J$ which comes from
applied spin current, $b_J^s=b_J-b_J^0$, $K_d$ corresponds integration of
dipole field. Here, $\alpha$ and $\gamma$ are renormalized parameter by the
same way with $n_0^\uda$. One can check that Eqs. (\ref{D10}) and (\ref{D11})
reproduces Eqs. (\ref{dw1 without SMF}) and (\ref{dw2 without SMF}) if
$\Delta n_s=0$ and $b_J^s=0$.

\subsection{1D DW motion in the presence of SMF and spin diffusion}
Now, we apply
\begin{eqnarray}
b_j^s&=&\frac{\mu_B\sigma }{eM_s}\frac{1-P^2}{2\lambda_{sf}}\int dx' e^{-|x-x'|/\lambda_{sf}}E_s(x',t),\\
E_s&=&\frac{\hbar}{2e\lambda}\sech^2\left(\frac{x-X}{\lambda}\right)\frac{\partial\phi}{\partial t},\\
\Delta \tilde{n}_s&=&\frac{\sigma\tau_{sf}}{e\lambda_{sf}^2}\frac{1-P^2}{1-P_n^2}\frac{k^2+P_nP\lambda_{sf}^{-2}}{k^2+\lambda_{sf}^{-2}}\frac{ik\tilde{E}_s}{k^2}.
\end{eqnarray}
to Eqs. (\ref{D10}) and (\ref{D11}).

First, we calculate the integral $\int \Delta
n_s\sech^2\left(\frac{x-X}{\lambda}\right)dx$ which corresponds to the effect
of spin density . Note that $\Delta n_s$ is an odd function of $x-X$. Hence,
the integrand is an odd function so the integral vanishes.

The next step is to calculate
$\int\frac{b_J^s}{\lambda}\sech^2\left(\frac{x-X}{\lambda}\right)\frac{dx}{2\lambda}$.
After some algebra,
\begin{eqnarray}
&&\int\frac{b_J^s}{\lambda}\sech^2\left(\frac{x-X}{\lambda}\right)\frac{dx}{2\lambda}\nonumber\\
&&=\frac{(1-P^2)\eta\zeta}{4\lambda^2}\frac{\partial\phi}{\partial t}\iint e^{-\zeta|u-u'|}\sech^2u\sech^2u'dudu'\nonumber\\
&&=\frac{(1-P^2)\eta\zeta}{4\lambda^2}\frac{\partial\phi}{\partial t}\int\frac{\pi\zeta k^2\csch^2\left(\frac{k\pi}{2}\right)}{k^2+\zeta^2}dk\nonumber\\
&&=\frac{(1-P^2)\eta\zeta}{4\lambda^2}\frac{\partial\phi}{\partial t}\int_0^\infty\frac{2\pi\zeta k^2\csch^2\left(\frac{k\pi}{2}\right)}{k^2+\zeta^2}dk
\end{eqnarray}
where $\zeta=\lambda/\lambda_{sf}$, $u=(x-X)/\lambda$, and
$u'=(x'-X)/\lambda$. Parseval's relation and convolution theorem of Fourier
transform is used at the third step. By using the identity
\begin{equation}
\csch^2\left(\frac{k\pi}{2}\right)=4\sum_{n=1}^\infty ne^{-n\pi k},
\end{equation}
the integral can be expressed by Laplace transform $\mathcal{L}$.
\begin{eqnarray}
\int_0^\infty\frac{2\pi\zeta k^2\csch^2\left(\frac{k\pi}{2}\right)}{k^2+\zeta^2}dk&=&\sum_{n=1}^\infty\mathcal{L}\left[\frac{8\pi\zeta nk^2}{k^2+\zeta^2}\right](n\pi)\nonumber\\
&=&\sum_{n=1}^\infty\int_0^\infty\frac{16}{n\pi}\frac{\sin n\pi\zeta u}{(u+1)^3}du.
\end{eqnarray}
Recall the Fourier series of a sawtooth function
\begin{equation}
\sum_{n=1}^\infty\frac{16}{n\pi}\sin n\pi x=8(2n+1-x),
\end{equation}
for $2n<x<2n+2$. Then the integral is given by
\begin{eqnarray}
&&\sum_{n=1}^\infty\int_0^\infty\frac{16}{n\pi}\frac{\sin n\pi\zeta u}{(u+1)^3}du\nonumber\\
&&=\sum_{n=0}^\infty\int_{\frac{2n}{\zeta}}^{\frac{2n+2}{\zeta}}\frac{8(2n+1-u\zeta)}{(1+u)^3}du\nonumber\\
&&=-4-4\zeta+2\zeta^2\sum_{n=0}^\infty\frac{1}{(n+\zeta/2)^2}.
\end{eqnarray}
By using the relation
\begin{equation}
\frac{d^n}{dz^n}\frac{\Gamma'(z)}{\Gamma(z)}=(-1)^{n+1}n!\sum_{k=1}^\infty\frac{1}{(z+k)^{n+1}},
\end{equation}
one obtains the integral in closed form as
\begin{eqnarray}
&&\frac{8}{3}\frac{F(\zeta)}{\zeta}\equiv\int_0^\infty\frac{2\pi\zeta k^2\csch^2\left(\frac{k\pi}{2}\right)}{k^2+\zeta^2}dk\nonumber\\
&&=-4-4\zeta+2\zeta^2\left.\left(\frac{\Gamma'(z)}{\Gamma(z)}\right)'\right|_{z=\frac{\zeta}{2}}\nonumber\\
&&=-4-4\zeta+2\zeta^2\frac{\Gamma''\left(\frac{\zeta}{2}\right)\Gamma\left(\frac{\zeta}{2}\right)-\Gamma'\left(\frac{\zeta}{2}\right)^2}{\left(\frac{\zeta}{2}\right)^2}.
\end{eqnarray}
The prefactor $8/3\zeta$ is to make
$\lim_{\lambda_{sf}\rightarrow0}F(\zeta)=1$. Therefore, we finally obtain
\begin{equation}
\int \frac{b_J^s}{\lambda}\sech^2\left(\frac{x-X}{2\lambda}\right)\frac{dx}{2\lambda}=(1-P^2)\frac{2\eta}{3\lambda^2}F(\zeta)\frac{\partial\phi}{\partial t},
\end{equation}
and, consequently,
\begin{eqnarray}
-\frac{\partial\phi}{\partial t}+\frac{\alpha}{\lambda}\frac{\partial X}{\partial t}&=&-\frac{\beta b_J^0}{\lambda}-\frac{2\beta\tilde{\eta}}{3\lambda^2}\frac{\partial\phi}{\partial t},\\
\frac{1}{\lambda}\frac{\partial X}{\partial t}+\alpha\frac{\partial\phi}{\partial t}&=&-\frac{b_J^0}{\lambda}-\frac{2\tilde{\eta}}{3\lambda^2}\frac{\partial\phi}{\partial t}-\frac{\gamma K_d}{M_s}\sin 2\phi,
\end{eqnarray}
where $\eta$ is renormalized parameter by spin diffusion defined as
$\tilde{\eta}=(1-P^2)F(\zeta)\eta$.
\end{appendix}

\end{document}